\documentclass{emulateapj} 
\usepackage{apjfonts,psfig,epsfig}
\begin{document} 

\slugcomment{AJ, in press}

\title{The Dwarf Spheroidal Companions to M31:  Variable Stars in Andromeda~I 
and Andromeda~III\footnotemark[1]} 
\footnotetext[1]{Based on observations with the NASA/ESA {\it Hubble Space Telescope}, 
obtained at the Space Telescope Science Institute, which is operated by the 
Association of Universities for Research in Astronomy, Inc., (AURA), under 
NASA Contract NAS 5-26555.}

\shortauthors{Pritzl et al.}
\shorttitle{And~I \& And~III Variable Stars}


\author{Barton J. Pritzl\altaffilmark{2} and Taft E. Armandroff}
\affil{National Optical Astronomy Observatory, P.O. Box 26732, Tucson, AZ
85726 \\email: pritzl@noao.edu, armand@noao.edu}
\author{George H. Jacoby}
\affil{WIYN Observatory, P.O. Box 26732, Tucson, AZ, 85726 \\email:
gjacoby@wiyn.org}
\author{G. S. Da Costa}
\affil{Research School of Astronomy and Astrophysics, Institute of
Advanced Studies, The Australian National University, Cotter Road, Weston,
ACT 2611, Australia \\email: gdc@mso.anu.edu.au}

\altaffiltext{2}{Current address:  Department of Physics and Astronomy, Macalester
College, 1600 Grand Ave., Saint Paul, MN 55105; pritzl@macalester.edu}

\begin{abstract}

We present the results of variable star searches of the M31 dwarf spheroidal 
companions Andromeda~I and Andromeda~III using the {\it Hubble Space 
Telescope}.  A total of 100 variable stars were found in Andromeda~I, while 
56 were found in Andromeda~III\@.  One variable found in Andromeda~I and 
another in Andromeda~III may be Population~II Cepheids.  In addition to this 
variable in Andromeda~III, another four variables are anomalous Cepheids.  
So far, no definite anomalous Cepheids have been discovered in Andromeda~I\@.  
We discuss the properties of these variables with respect to those found in the 
other dwarf spheroidal galaxies and revisit the anomalous Cepheid 
period-luminosity relations.  We found 72 fundamental mode RR~Lyrae stars 
and 26 first-overtone mode RR~Lyrae stars in Andromeda~I giving mean periods 
of 0.575~day and 0.388~day, respectively.  One likely RR~Lyrae in Andromeda~I 
remains unclassified due to a lack of F555W data.  For Andromeda~III, 39 RR~Lyrae 
stars are pulsating in the fundamental mode with a mean period of 0.657~day 
and 12 are in the first-overtone mode with a mean period of 0.402~day.  
Using the mean metal abundances derived from the red giant branch colors, 
the mean RRab period for Andromeda~I is consistent with the mean 
period - metallicity
relation seen in the RR~Lyrae populations of Galactic globular clusters, while 
Andromeda~III is not, having too large a mean RRab period for its abundance.  In 
Andromeda~I, we found two RR~Lyrae stars which are noticeably fainter than 
the horizontal branch.  We discuss the possibility that these stars are 
associated with the recently discovered stellar stream in the halo of 
M31.  

Using various methods, we estimate the mean metallicity of the RR~Lyrae 
stars to be $\langle {\rm [Fe/H]}\rangle \approx -1.5$ for Andromeda~I and 
$\approx -1.8$ for Andromeda~III\@.  These estimates match well with other mean 
metallicity estimates for the galaxies.  Interestingly, a comparison of the 
period-amplitude diagrams for these two galaxies with other dwarf 
spheroidal galaxies shows that Andromeda~III is lacking in shorter period, 
higher amplitude RR~Lyrae stars.  This may be a consequence of the very
red horizontal branch morphology in this dSph.
Not including the two faint RR~Lyrae stars, we find 
$\langle V_{\rm RR} \rangle = 25.14\pm0.04$~mag for Andromeda~I resulting 
in a distance of $765\pm25$~kpc.  For Andromeda~III, 
$\langle V_{\rm RR} \rangle = 25.01\pm0.04$~mag giving a distance of 
$740\pm20$~kpc.  These distance estimates are consistent with those 
previously found for these galaxies.  We discuss the relation between 
the specific frequency of the anomalous Cepheids in dwarf spheroidal 
galaxies and the mean metallicity of the galaxy, finding that the M31 
dwarf spheroidal galaxies follow the same relations as the Galactic 
dwarf spheroidal galaxies.  We also find that the period-luminosity 
relations of anomalous Cepheids and short-period Cepheids are distinct, 
with the short-period Cepheids having higher luminosities at a given period.  

\end{abstract}


\section{Introduction} 

\begin{figure*}[t]
 \centerline{\psfig{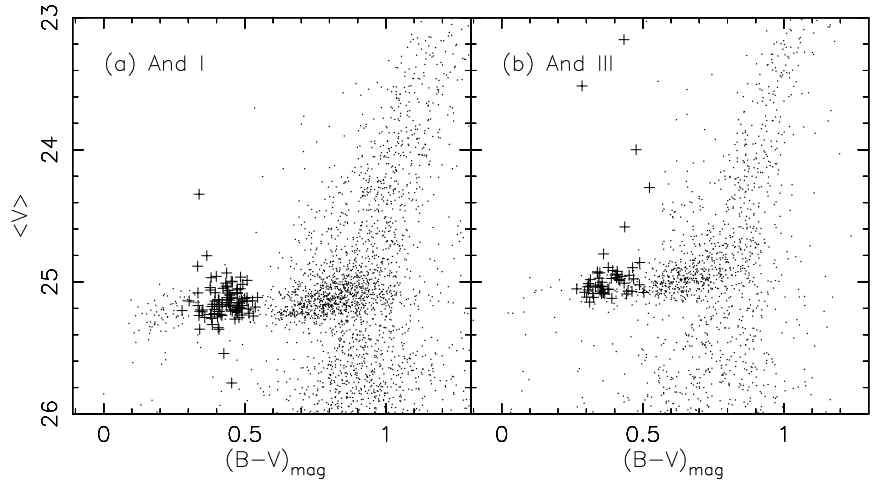}}
 \caption{And~I (a) and And~III (b) color-magnitude diagrams using the
  photometry of DACS00 and DAC02 except for the
  variable stars identified in this work which are plotted as plus
  signs.  The anomalous Cepheids or Population~II Cepheids occur above
  the level of the horizontal branch.  In addition, we note the presence
  of two fainter RR~Lyrae in the And~I color-magnitude diagram.}
 \label{Fig01}
\end{figure*}

\begin{figure*}[t]
  \centerline{\psfig{figure=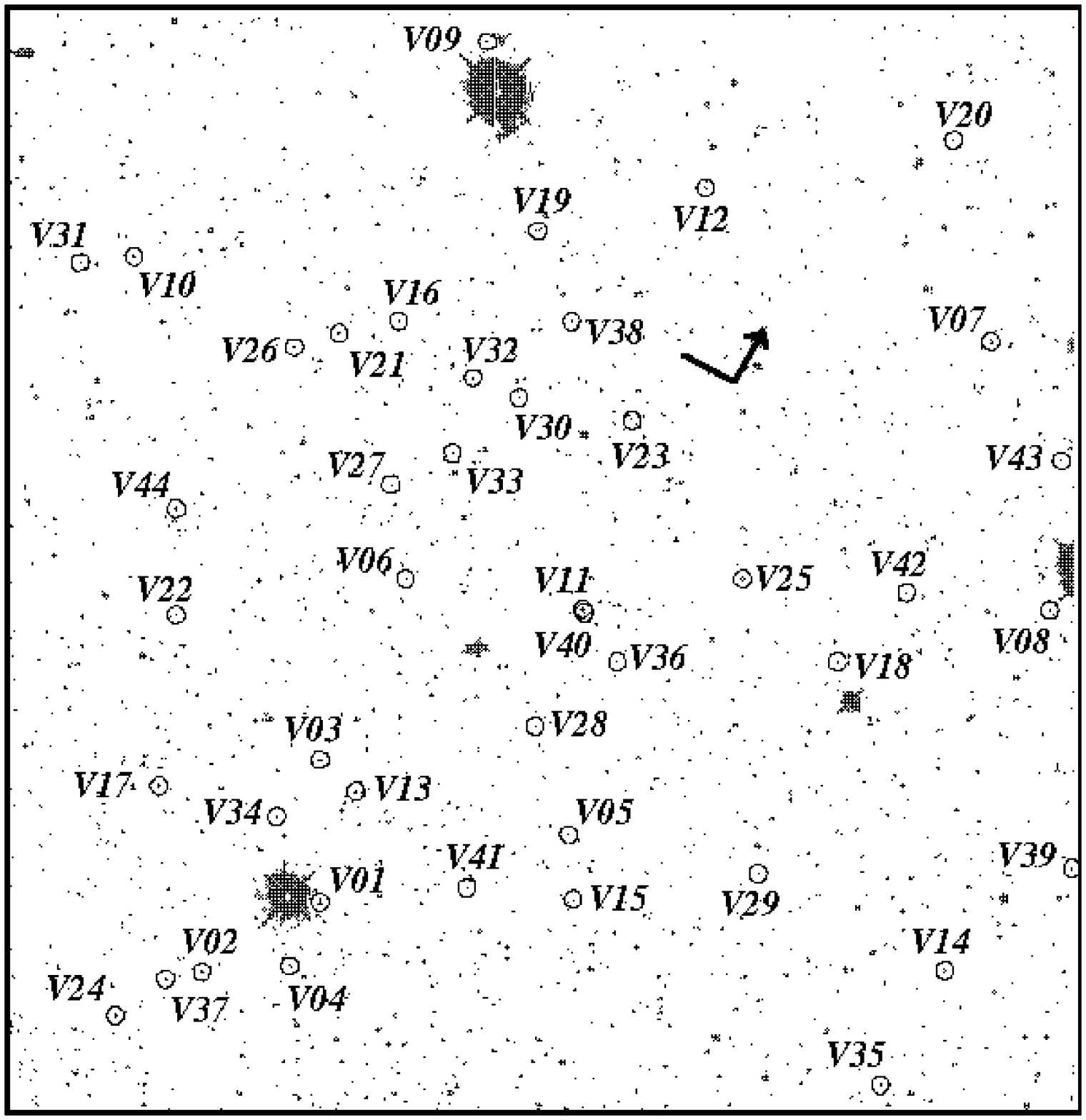,height=3.25in,width=3.50in}}
  \caption{Finding charts for the And~I variable stars.  The WFC2
   ($1.2\arcmin$x$1.3\arcmin$), WFC3 ($1.3\arcmin$x$1.2\arcmin$),
   and WFC4 ($1.2\arcmin$x$1.2\arcmin$) images are each shown in a
   panel.  North and east directions are shown with the arrow pointing
   toward the north.}
  \label{Fig02a}
\end{figure*}

\begin{figure*}[t]
  \figurenum{2 cont}
  \centerline{\psfig{figure=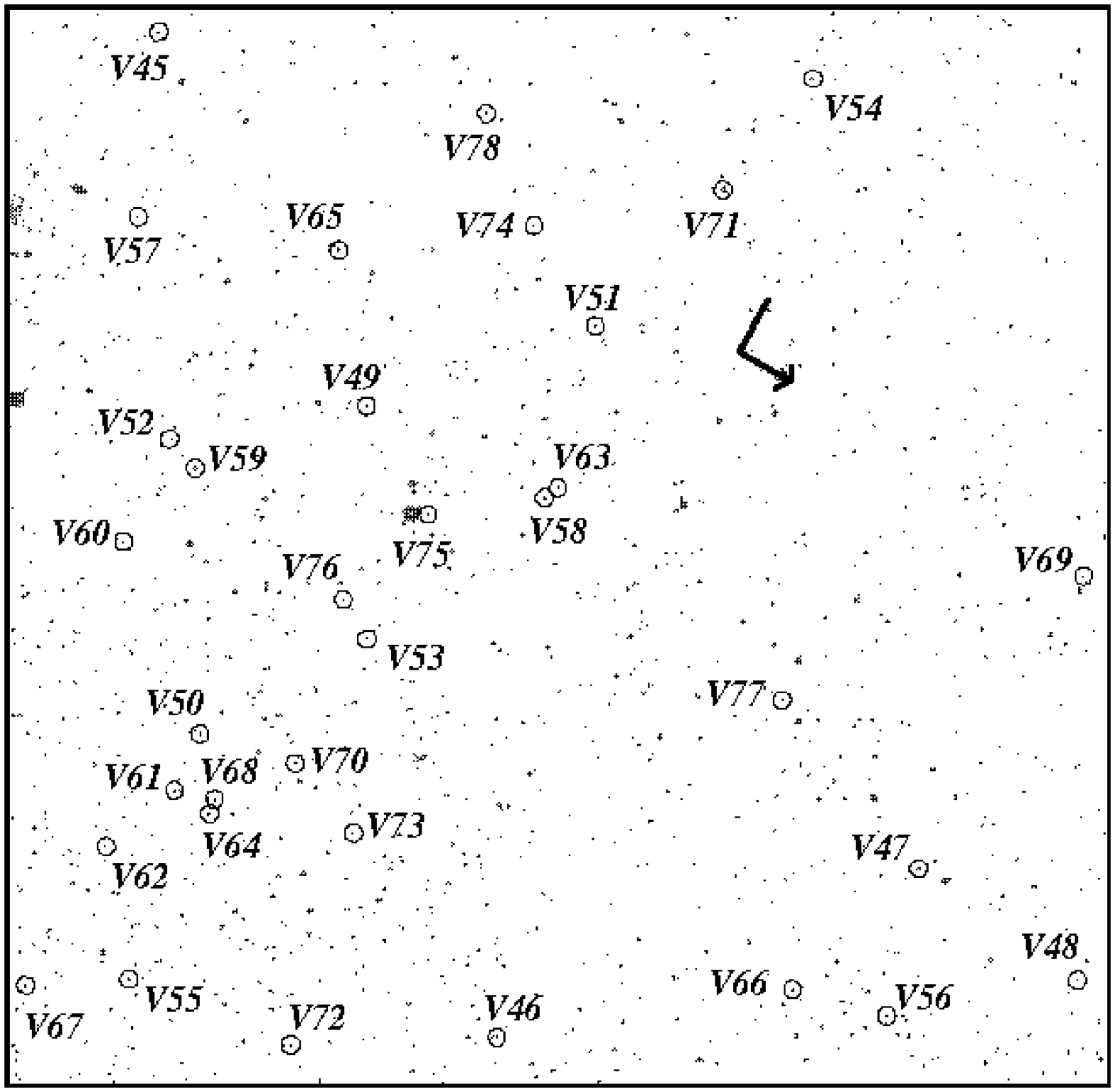,height=3.25in,width=3.50in}}
  \caption{Finding charts for the And~I variable stars.  The WFC2
   ($1.2\arcmin$x$1.3\arcmin$), WFC3 ($1.3\arcmin$x$1.2\arcmin$),
   and WFC4 ($1.2\arcmin$x$1.2\arcmin$) images are each shown in a
   panel.  North and east directions are shown with the arrow pointing
   toward the north.}
  \label{Fig02b}
\end{figure*}

\begin{figure*}[t]
  \figurenum{2 cont}
  \centerline{\psfig{figure=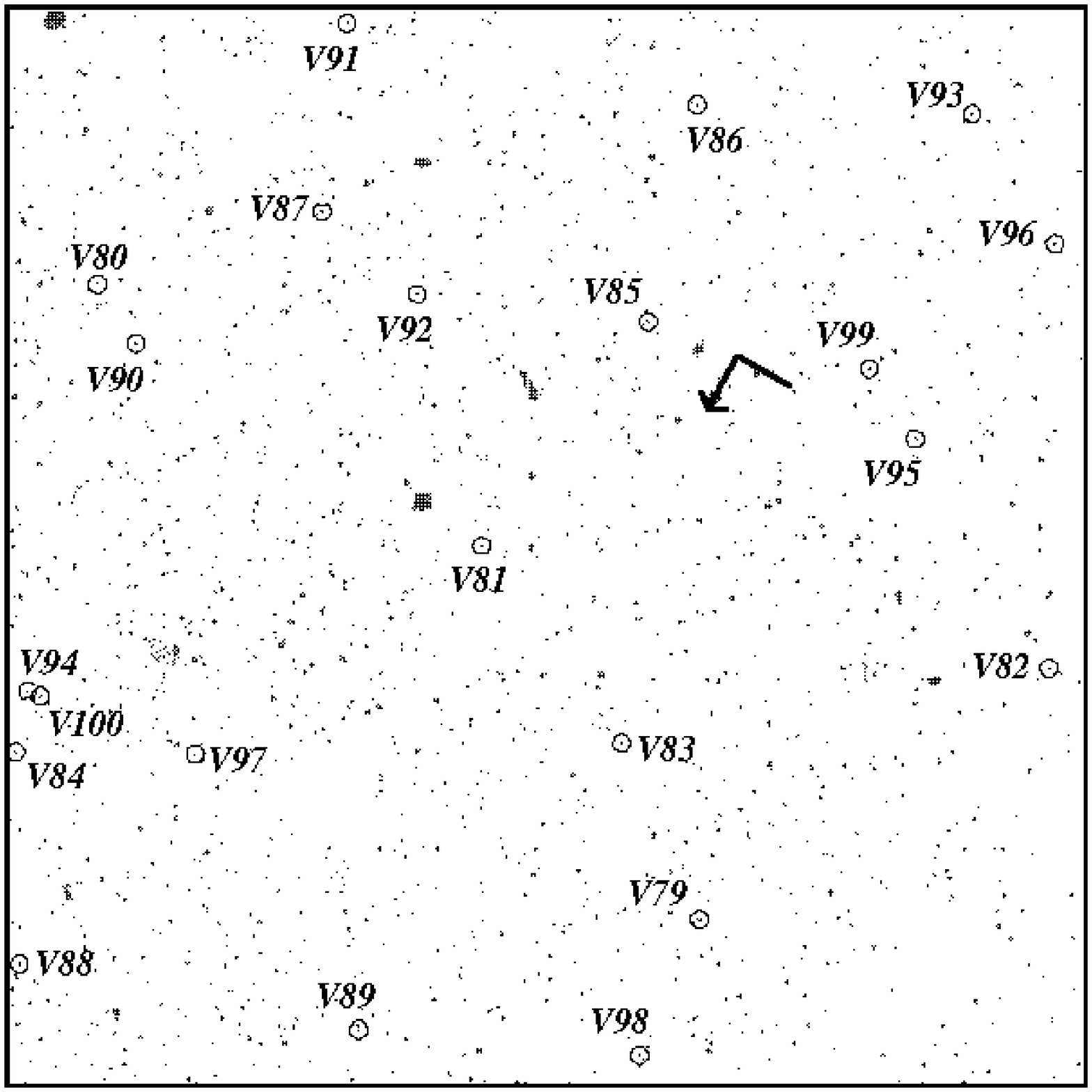,height=3.25in,width=3.50in}}
  \caption{Finding charts for the And~I variable stars.  The WFC2
   ($1.2\arcmin$x$1.3\arcmin$), WFC3 ($1.3\arcmin$x$1.2\arcmin$),
   and WFC4 ($1.2\arcmin$x$1.2\arcmin$) images are each shown in a
   panel.  North and east directions are shown with the arrow pointing
   toward the north.}
  \label{Fig02c}
\end{figure*}

\begin{figure*}[t]
  \centerline{\psfig{figure=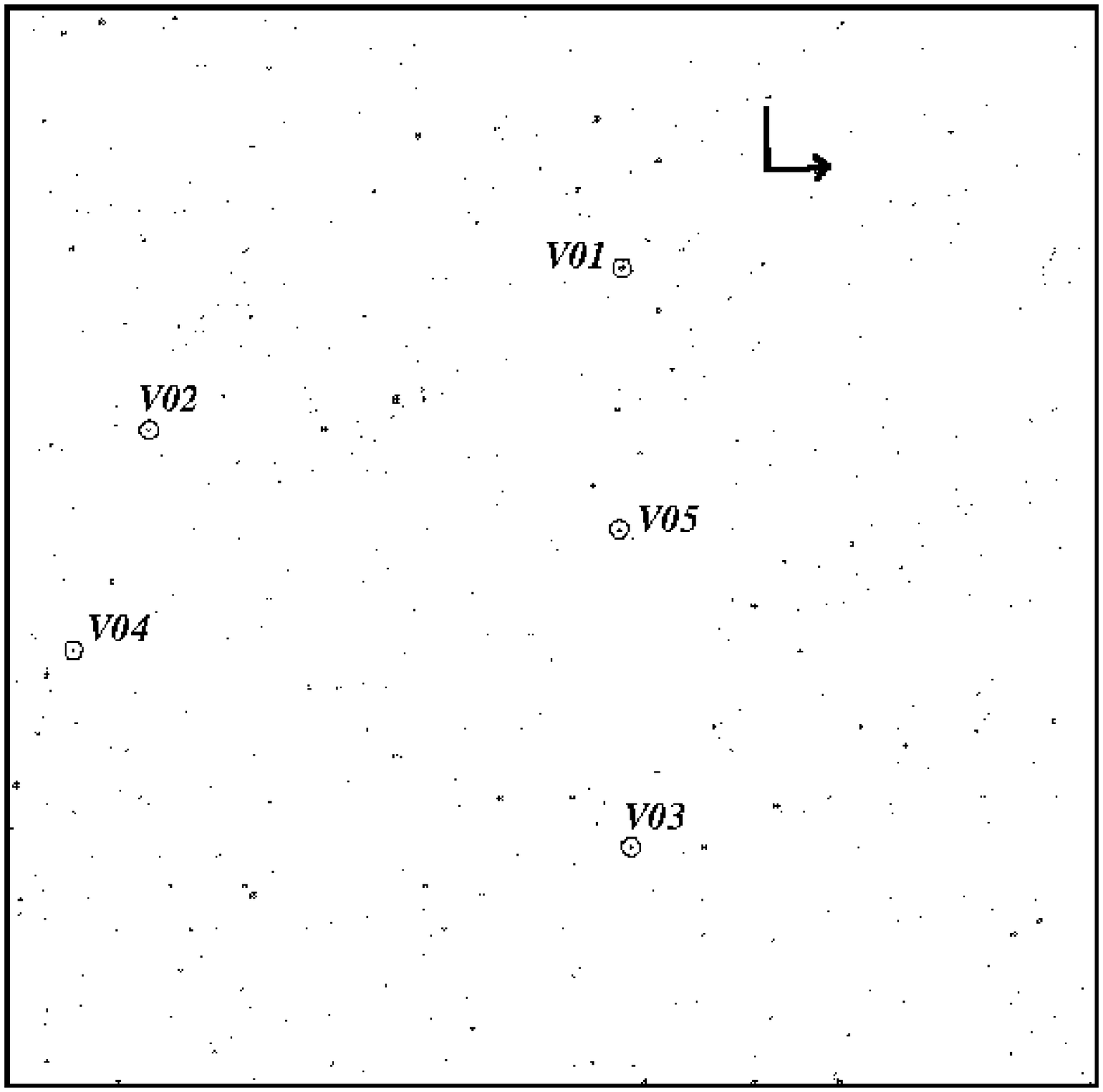,height=3.35in,width=3.50in}}
  \caption{Finding charts for the And~III variable stars.  The PC 
   ($0.5\arcmin$x$0.5\arcmin$), WFC2
   ($1.2\arcmin$x$1.3\arcmin$), WFC3 ($1.3\arcmin$x$1.2\arcmin$),
   and WFC4 ($1.2\arcmin$x$1.2\arcmin$) images are each shown in a
   panel.  North and east directions are shown with the arrow pointing
   toward the north.}
  \label{Fig03a}
\end{figure*}

\begin{figure*}[t]
  \figurenum{3 cont}
  \centerline{\psfig{figure=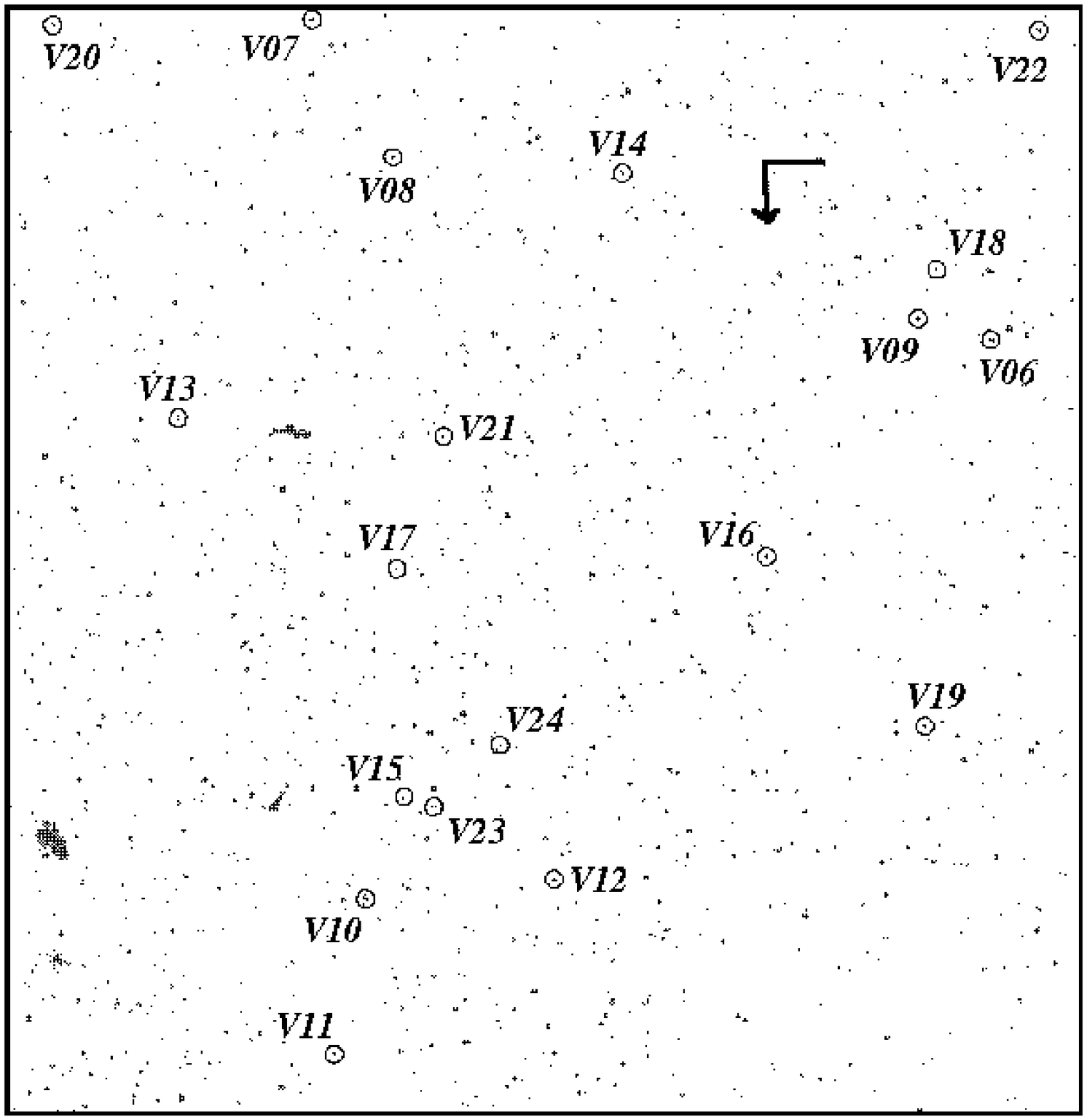,height=3.35in,width=3.50in}}
  \caption{Finding charts for the And~III variable stars.  The PC 
   ($0.5\arcmin$x$0.5\arcmin$), WFC2
   ($1.2\arcmin$x$1.3\arcmin$), WFC3 ($1.3\arcmin$x$1.2\arcmin$),
   and WFC4 ($1.2\arcmin$x$1.2\arcmin$) images are each shown in a
   panel.  North and east directions are shown with the arrow pointing
   toward the north.}
  \label{Fig03b}
\end{figure*}

\begin{figure*}[t]
  \figurenum{3 cont}
  \centerline{\psfig{figure=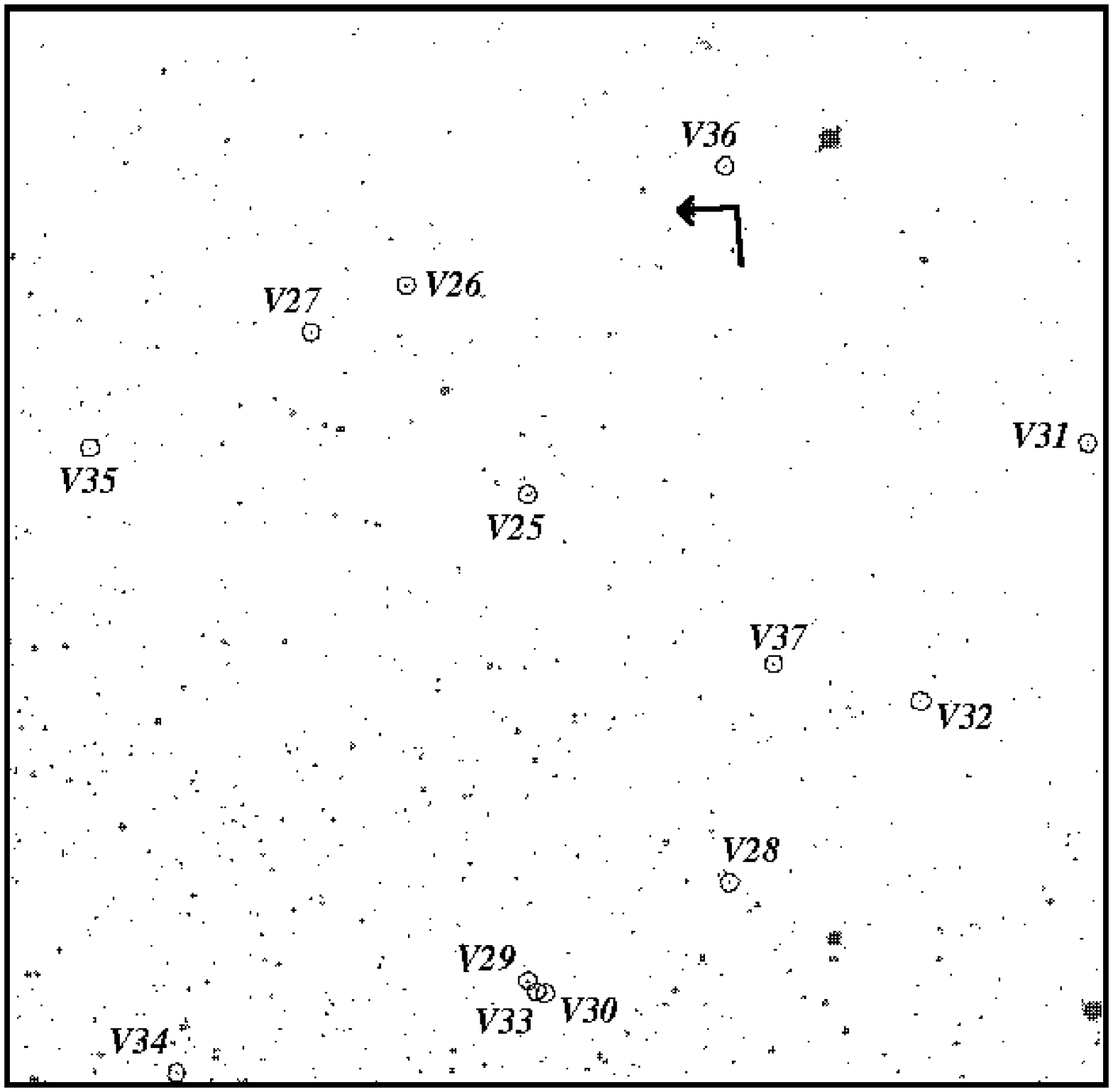,height=3.25in,width=3.50in}}
  \caption{Finding charts for the And~III variable stars.  The PC, 
   ($0.5\arcmin$x$0.5\arcmin$), WFC2
   ($1.2\arcmin$x$1.3\arcmin$), WFC3 ($1.3\arcmin$x$1.2\arcmin$),
   and WFC4 ($1.2\arcmin$x$1.2\arcmin$) images are each shown in a
   panel.  North and east directions are shown with the arrow pointing
   toward the north.}
  \label{Fig03c}
\end{figure*}

\begin{figure*}[t]
  \figurenum{3 cont}
  \centerline{\psfig{figure=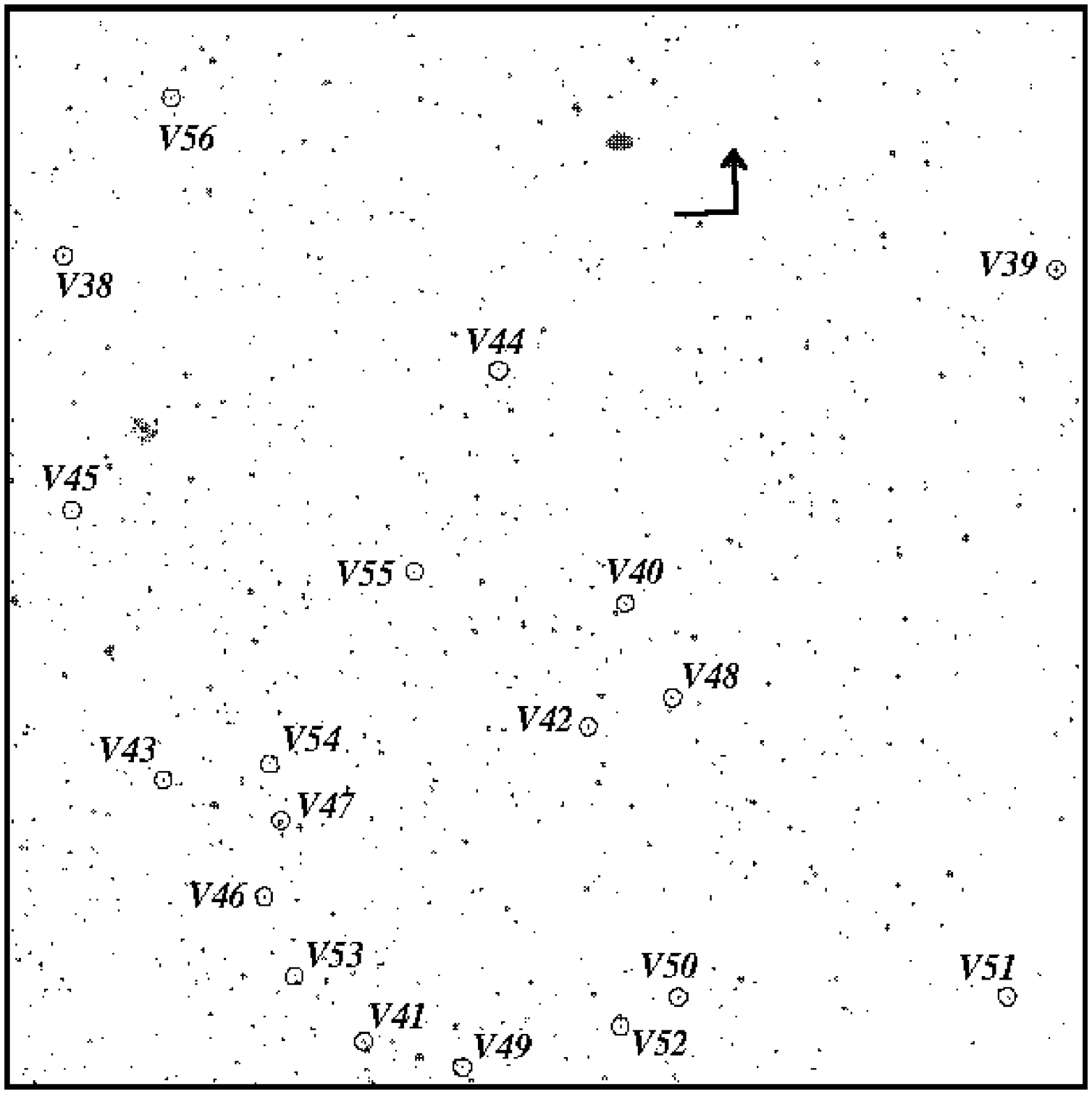,height=3.25in,width=3.50in}}
  \caption{Finding charts for the And~III variable stars.  The PC 
   ($0.5\arcmin$x$0.5\arcmin$), WFC2
   ($1.2\arcmin$x$1.3\arcmin$), WFC3 ($1.3\arcmin$x$1.2\arcmin$),
   and WFC4 ($1.2\arcmin$x$1.2\arcmin$) images are each shown in a
   panel.  North and east directions are shown with the arrow pointing
   toward the north.}
  \label{Fig03d}
\end{figure*}

The Galactic dwarf spheroidal (dSph) galaxies are known to have a diversity 
of star formation histories, from mainly old (age $>$ 10~Gyr) stellar populations such 
as Ursa Minor (e.g., Olszewski \& Aaronson 1985), to Fornax which had star 
formation as late as 0.3-0.4 Gyrs ago (Saviane, Held, \& Bertelli 2000).  
While there have been many studies of the Galactic dSph galaxies, until 
recently little was known about the stellar populations of the M31 dSph 
galaxies.  In fact, it was only in the past few years that half of the 
known M31 dSph galaxies were discovered (Andromeda~V:  Armandroff, Davies, 
\& Jacoby 1998; Andromeda~VI [Pegasus]:  Armandroff, Jacoby, \& 
Davies 1999, Karachentsev \& Karachentseva 1999; Andromeda~VII [Cassiopeia]:  
Karachentsev \& Karachentseva 1999).  

A series of observations of the M31 dSph galaxies with the {\it Hubble 
Space Telescope} (HST) Wide Field Planetary Camera~2 (WFPC2) has been used 
to investigate their stellar populations to levels fainter than the 
horizontal branch (HB).  We are also able to use these observations to 
investigate the variable star content of these dSph galaxies.  
The variable star content is important because it provides 
insight to the age, metallicity, and distance to the galaxy in which the variables 
are found.  We have already discussed the variable stars in Andromeda~VI 
(And~VI; Pritzl et al.\ 2002; henceforth Paper~I) and Andromeda~II (And~II; 
Pritzl et al.\ 2004; henceforth Paper~II)\@.  In this third paper of the 
series, we investigate the variable stars in Andromeda~I (And~I) and 
Andromeda~III (And~III).  

Of the M31 dSph companions, And~I lies closest to M31.  It is also on the 
high end of the metallicities for dSph galaxies ($\langle{\rm [Fe/H]}\rangle 
= -1.46\pm0.12$; Da~Costa et al.\ 2000, henceforth DACS00).  The 
color-magnitude diagram (CMD) shows the typical properties seen in other 
dSph galaxies such as a predominantly red HB and an internal abundance 
spread.  A radial gradient in the HB stars was also detected where there 
are relatively more blue HB stars beyond the core radius of the galaxy 
(Da~Costa et al.\ 1996; henceforth DACS96).

And~III also lies in the outer halo of M31 and is the next closest dSph to M31 
after And~I.  The HB in the CMD for this dSph galaxy is significantly 
redder than that seen in the other M31 dSph galaxies that have been observed.  
And~III is also one of the most metal-poor 
of the M31 dSph companions ($\langle{\rm [Fe/H]}\rangle = -1.88\pm0.11$; 
Da~Costa, Armandroff, \& Caldwell 2002, henceforth DAC02).  Further, while 
there is 
also an internal abundance spread in And~III, it is low compared to most other 
dSph galaxies.

The variable star content of And~I was initially investigated in DACS96 
and later in more detail in Da~Costa et al.\ 
(1997), while that for And~III was investigated in DAC02.  In this paper we 
expand upon those works bringing to bear the techniques described below and 
in Paper~I\@.  After describing the processing of the data, we discuss 
the variable star contents of these galaxies and compare them with those for 
other dSph galaxies.  We also discuss how the variable star 
populations reflect the properties of And~I and And~III.

\section{Observations and Reductions} 

Observations of And~I were taken by the HST/WFPC2 as part of the GO Program 
5325 on 1994 August 11 and 16, using the same orientation.  
Three 1800~s integrations through the F555W filter and six 1800~s 
integrations through the F450W filter were taken in the first set of 
observations.  Identical exposure times were used for the second set of 
observations, but there were four F555W and six F450W integrations.  

For And~III, the observations were taken for the GO Program 7500 on 1999 
February 22 and 26 with the same orientation.  Each set of observations 
had four observations of 1200~s in F555W and eight observations of 1300s 
in F450W\@.  As for And~I, the two sets of observations were slightly offset 
from each other to minimize the impact of instrumental defects.  For 
And~III, the 
field center was offset from the center of the dSph galaxy to avoid 
bright foreground stars (see Fig.~1 of DAC02). 

The point-spread function fitting photometry was performed with {\sc allframe} 
(Stetson 1994) as described in Paper~I\@.  Similarly, the aperture and 
charge-transfer efficiency corrections along with the $B$,$V$ calibrations 
were carried out in the same fashion.  As with And~II, we have the 
photometry used in creating the color-magnitude diagrams (CMDs) so we may 
compare our photometry with that of DACS00 and DAC02.  
We list in Table~1 the mean differences between the two sets of photometry.
There are obvious systematic differences that are of 
similar size and sign as those found for And~II in Paper~II\@.  The origin 
of these systematic differences is unclear, although it likely is due to the 
different techniques and calibration processes.  Fortunately they appear 
to be just zero point offsets:  there is no dependance of these 
differences on magnitude.  We have applied them to our data in order to 
keep the present photometry consistent with the earlier work.  It should 
also be noted that DAC02 found their HST $V$-band photometry to be consistent 
with ground-based observations of And~III.

\begin{figure*}[t]
  \centerline{\psfig{figure=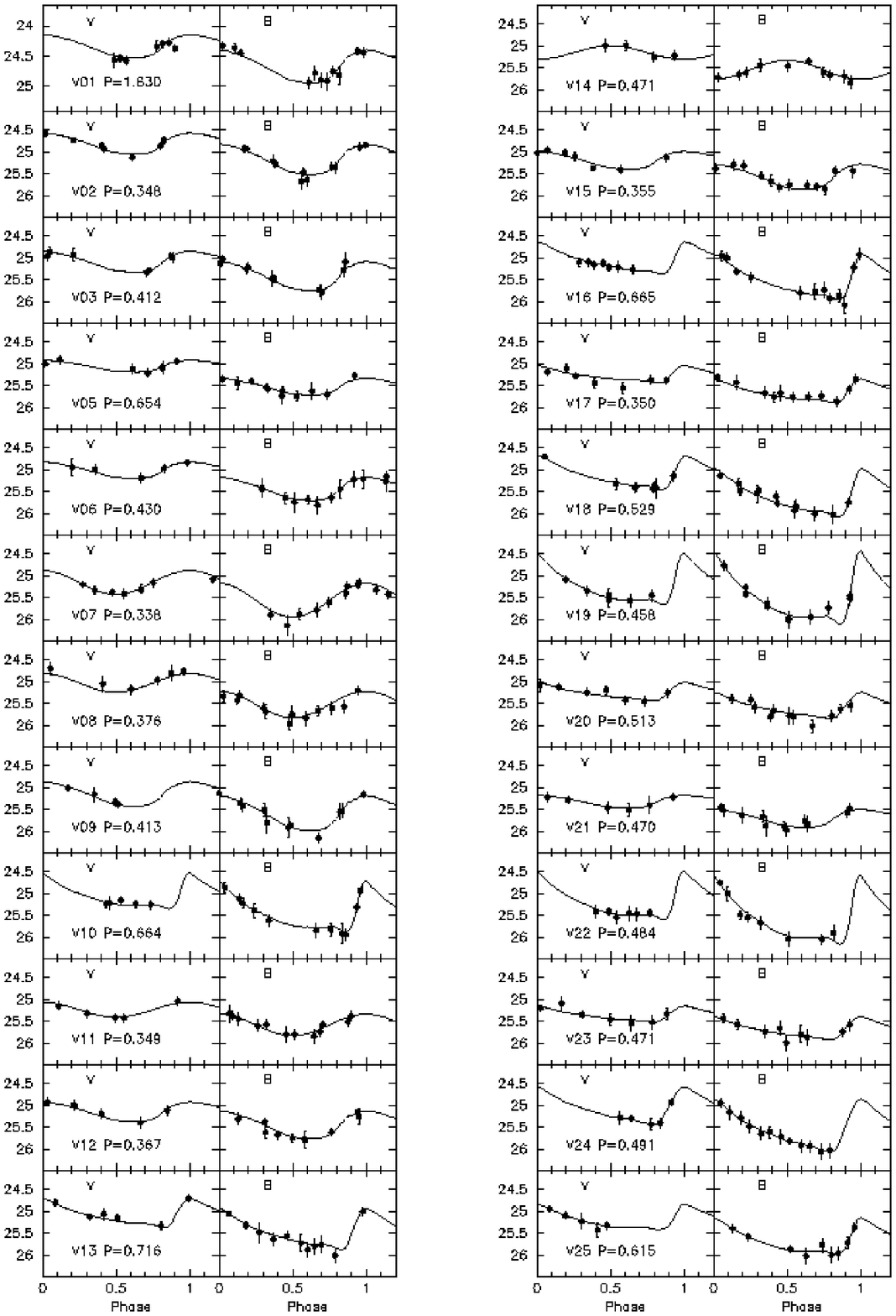,height=7.00in,width=5.00in}}
  \caption{And~I variable star light curves.  The observations are shown
   as filled circles and the fitted templates are displayed as curves.}
  \label{Fig04}
\end{figure*}

\begin{figure*}[t]
  \figurenum{4 cont}
  \centerline{\psfig{figure=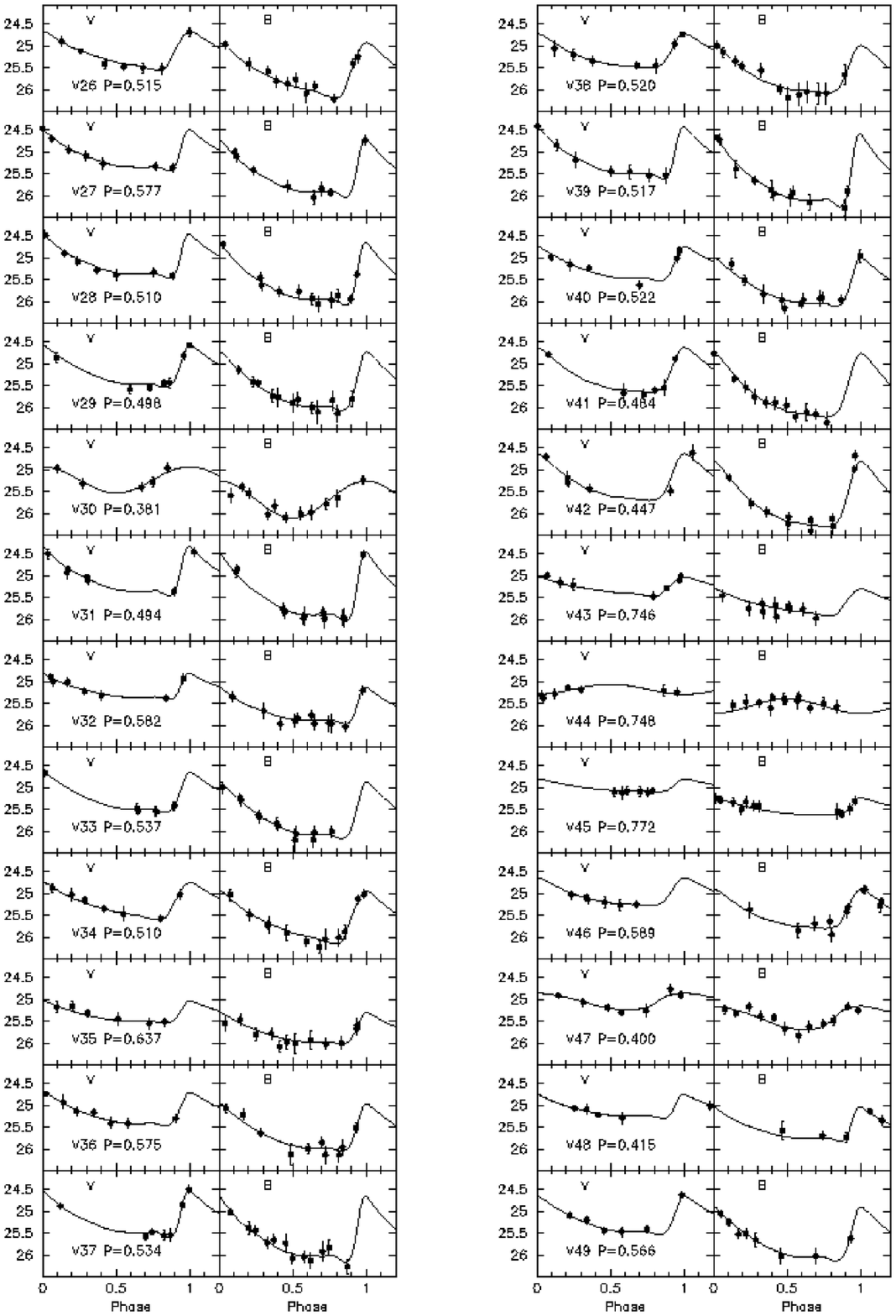,height=7.00in,width=5.00in}}
  \caption{And~I variable star light curves.  The observations are shown
   as filled circles and the fitted templates are displayed as curves.}
  \label{Fig04}
\end{figure*}

\begin{figure*}[t]
  \figurenum{4 cont}
  \centerline{\psfig{figure=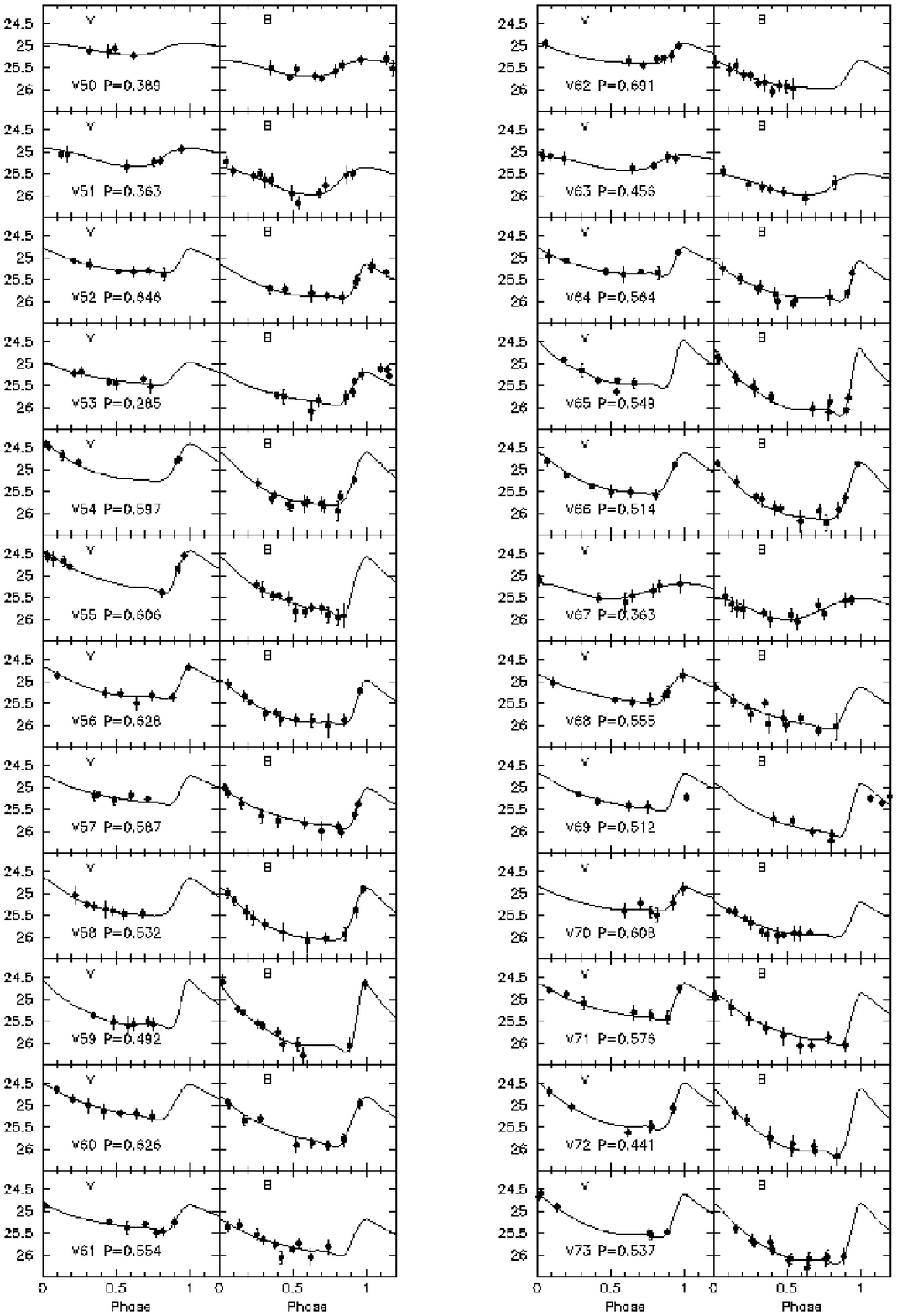,height=7.00in,width=5.00in}}
  \caption{And~I variable star light curves.  The observations are shown
   as filled circles and the fitted templates are displayed as curves.}
  \label{Fig04}
\end{figure*}

\begin{figure*}[t]
  \figurenum{4 cont}
  \centerline{\psfig{figure=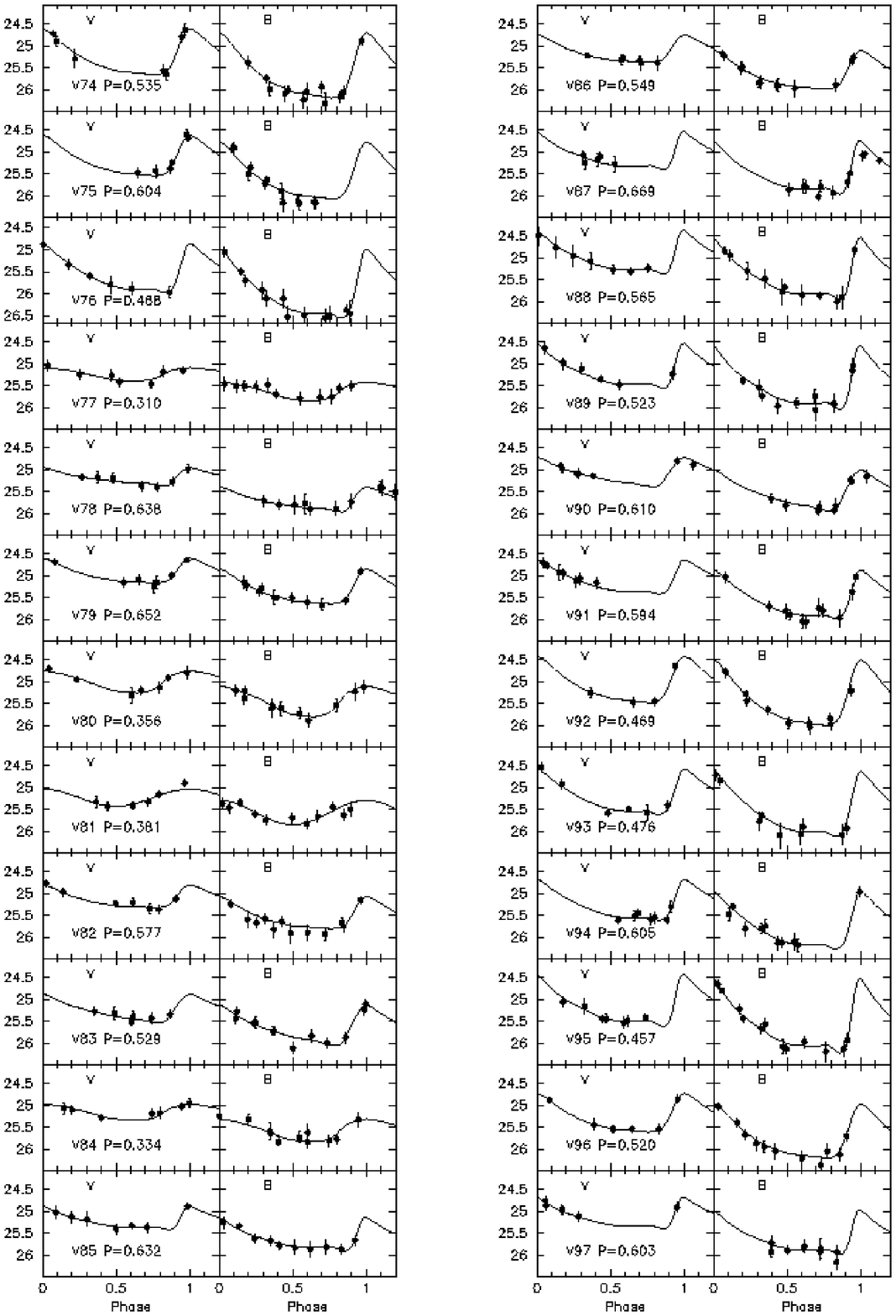,height=7.20in,width=5.00in}}
  \caption{And~I variable star light curves.  The observations are shown
   as filled circles and the fitted templates are displayed as curves.}
  \label{Fig04}
\end{figure*}

\begin{figure*}[t]
  \figurenum{4 cont}
  \centerline{\psfig{figure=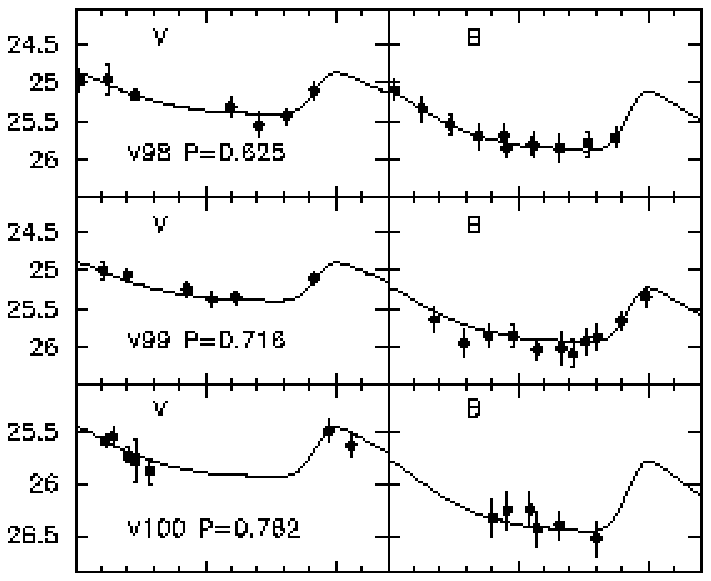,height=1.75in,width=2.50in}}
  \caption{And~I variable star light curves.  The observations are shown
   as filled circles and the fitted templates are displayed as curves.}
  \label{Fig04}
\end{figure*}

\begin{figure*}[t]
  \centerline{\psfig{figure=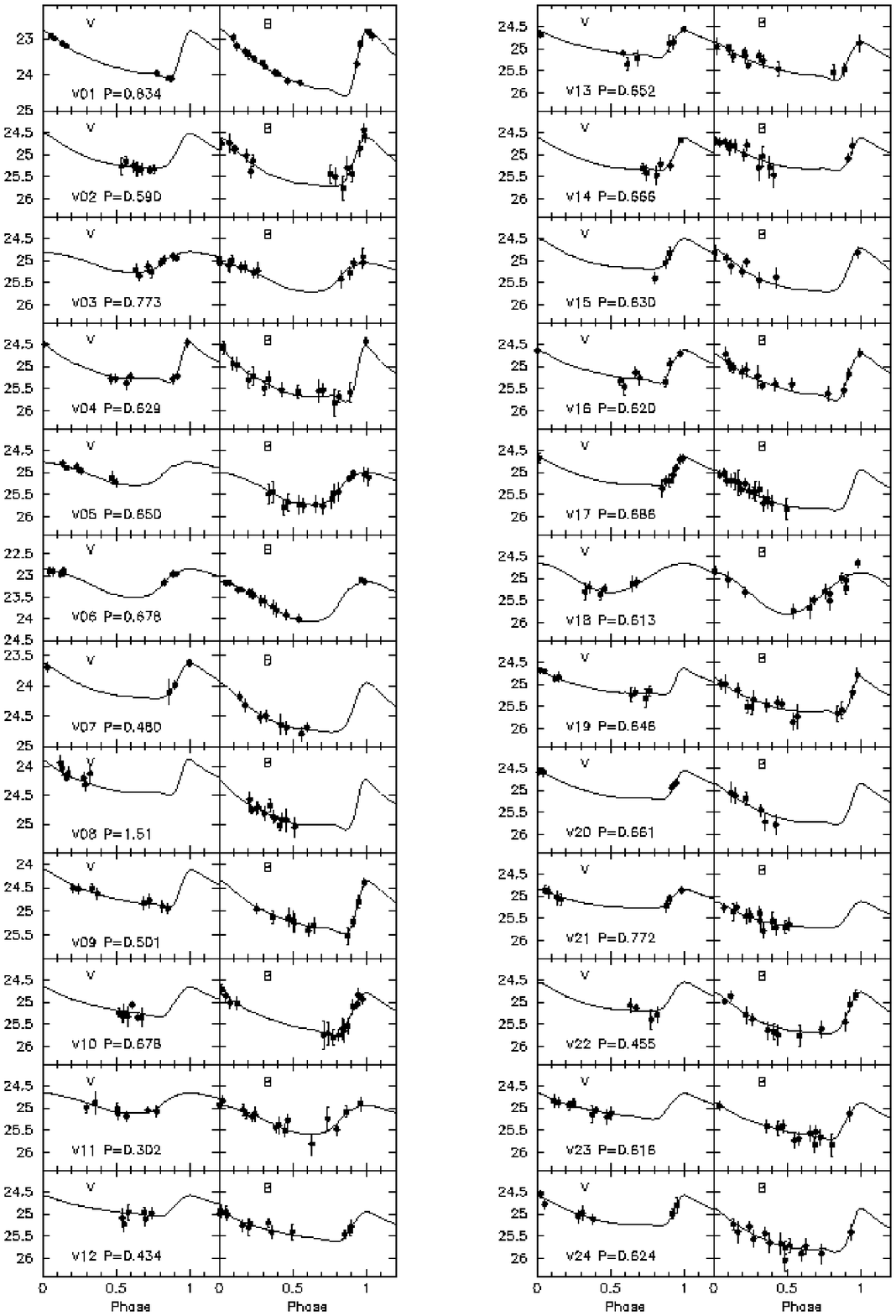,height=7.00in,width=5.00in}}
  \caption{And~III variable star light curves.  The observations are shown
   as filled circles and the fitted templates are displayed as curves.}
  \label{Fig05}
\end{figure*}

\begin{figure*}[t]
  \figurenum{5 cont}
  \centerline{\psfig{figure=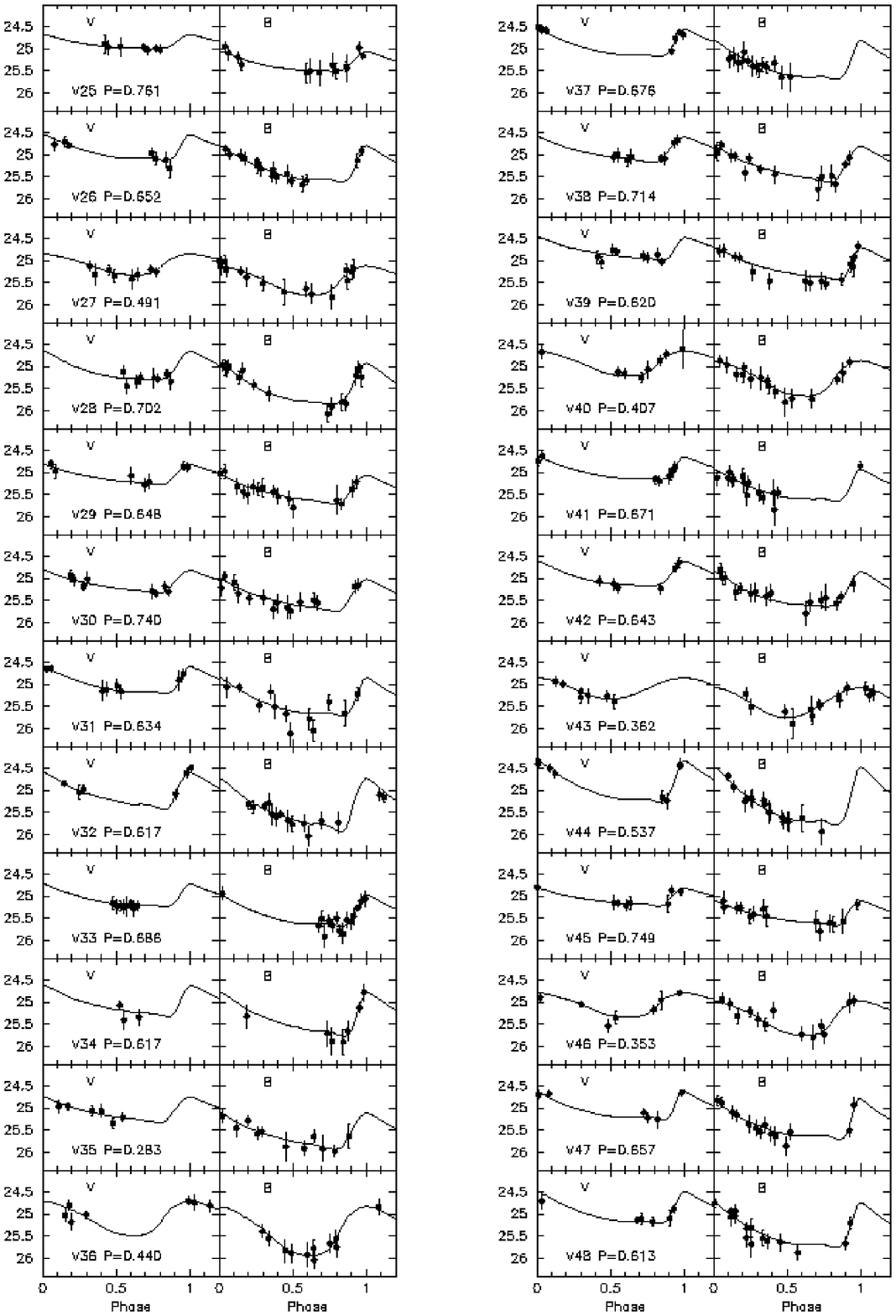,height=7.00in,width=5.00in}}
  \caption{And~III variable star light curves.  The observations are shown
   as filled circles and the fitted templates are displayed as curves.}
  \label{Fig05}
\end{figure*}

\begin{figure*}[t]
  \figurenum{5 cont}
  \centerline{\psfig{figure=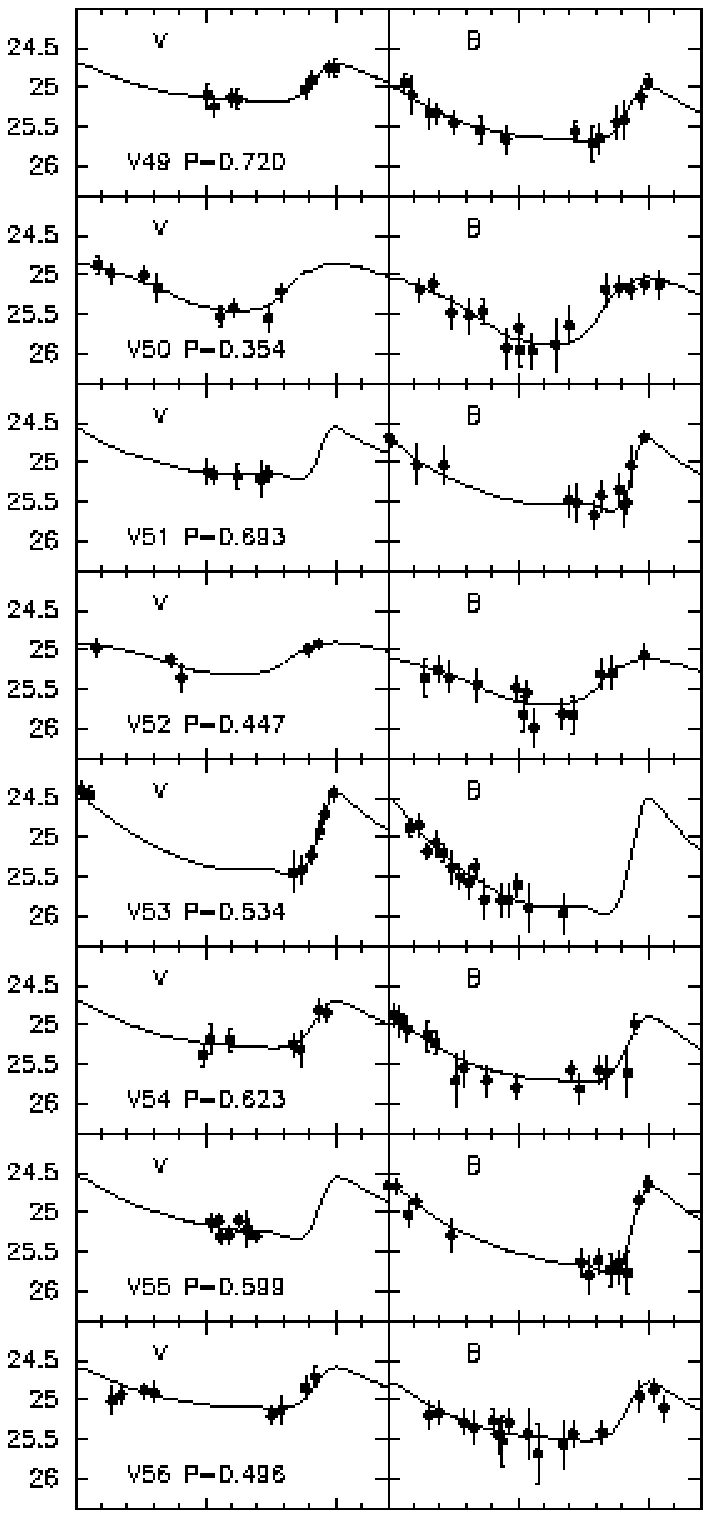,height=4.50in,width=2.50in}}
  \caption{And~III variable star light curves.  The observations are shown
   as filled circles and the fitted templates are displayed as curves.}
  \label{Fig05}
\end{figure*}

\section{Variable Stars} 

As was done in Papers~I and II, the photometry of the individual images was 
searched for variable stars using {\sc daomaster}, a routine created by 
P. B. Stetson.  {\sc daomaster} compares the rms scatter in the photometric 
values with that expected from the photometric errors returned by the 
{\sc allframe} program.  For And~I, we do not include the PC data since only a 
small number of RR~Lyrae (RRL) stars were found and no candidate anomalous 
Cepheids (ACs).  On the other hand, we did find one AC on the PC for And~III.  
Therefore we include the RRL stars found on that chip in our analysis.  
Period searches of the variable stars were done using the routines 
created by A. C. Layden as outlined in Papers~I and II.  The routines determine 
the most likely period from the $\chi^2$ minima by fitting the photometry of the 
variable star with 10 templates over a selected range of periods.  A cubic spline 
was fitted to each template light curve to determine the mean magnitudes and colors.  
The amplitudes derive from the template fits.

As discussed in Paper~I, the nature of the observations makes the determination of 
precise periods and magnitudes a challenge.  The data for both filters are 
combined to allow for more accurate magnitudes and periods (see Paper~I for 
more details).  As was done in the previous papers in this series, we used the 
period-amplitude diagram to test the periods of the RRL stars.  When one of these 
stars was scattered from the majority in the period-amplitude daigram we revised the 
period to reduce the scatter.  Although there is a slight possibility that a 
small number of And~I and And~III RRL stars may have an alias period, the combination of the 
template-fitting program and the period-amplitude diagram reduces the likelihood 
of this happening.

Figure~1 shows the CMDs of DACS00 (And~I) and DAC02 (And~III) in which we have 
replaced their photometry with that derived here for all the stars found to be 
variable in our survey.  Further, we checked all the DACS00 and DAC02 stars in 
Figure~1 that fall in the region of the instability strip along the HB or brighter 
for variability.  However, no additional variables were found.  Each variable is plotted 
according to its intensity-weighted $V$ magnitude and magnitude-weighted 
$(\bv)$ color.  Table~2 lists the photometric data for the variable stars 
in And~I, while Table~3 lists these for And~III.  For both tables, 
column~1 lists the star's ID, while the next two columns give the RA and 
Dec.  Column~4 lists the period of each star.  The intensity-weighted 
$\langle V \rangle$ and $\langle B \rangle$ magnitudes along with the 
magnitude-weighted colors $(\bv)_{\rm mag}$ are shown in columns 5 - 7.  
Columns~8 and 9 give the $V$ and $B$ amplitudes of the variable stars.  
The classifications of the variable stars are listed in column~12.  The 
remaining columns will be discussed later in the paper.  Figures~2 and 3 
are the finding charts for And~I and And~III, respectively, while  
Tables~4 - 7 give the photometric $B$ and $V$ data for And~I and And~III\@.  
The light curves for the variable stars in And~I are shown in Figure~4, 
while those for And~III are in Figure~5.

For And~I, there is one clear supra-HB variable star in the CMD (Fig.\ 1a).  
This 
star may be a Population~II Cepheid (P2C) and is discussed in the following 
section.  There are also two variable stars that are 
clearly fainter than the HB\@.  These stars will be discussed in \S5.1.1.

The And~III CMD in Fig.~1b shows five variable stars brighter than 
the RRL along the HB\@.  DAC02 discovered two of these stars and suggested 
one (V08$=$WF2-1398 in DAC02) may be a P2C\@.  This possibility is also 
discussed in the next section.  The two CMDs also show that And~III, 
despite having a lower mean metallicity than And~I, has a redder HB 
(cf.\ DAC02).

\section{Anomalous Cepheids and Population~II Cepheids} 

As mentioned above, And~I has one supra-HB variable star and And~III has five.  
Following the prescription used in Paper~I, we list the absolute magnitudes 
for these stars in Table~8 assuming a distance modulus and visual absorption of 
$24.49\pm0.06$~mag and $0.16\pm0.03$ for And~I (DACS00) and $24.38\pm0.06$~mag 
and $0.17\pm0.03$ for And~III (DAC02).  The period-luminosity relation shown
in Figure~6 plots the ACs from Paper~I and Paper~II along with the supra-HB 
variable stars in And~I and And~III\@.  
Four of the supra-HB variable stars in And~III fit in well with the other ACs
in this Figure confirming their classification as ACs.  These ACs are also 
found to be consistent with the other ACs in a period-amplitude diagram.  

In addition to the AC period-luminosity (P-L) relation lines, we plot 
in Figure~6 the P2C P-L relation for $P<10$~days using equation~11 in McNamara 
(1995).  Clearly, V01 in And~I and V08 in And~III fall near the P2C 
line for their given magnitude and period.  The way the data were taken 
and the relatively long apparent periods for these stars combine to 
generate significant gaps in their light curves.  
This leads to higher uncertainty in their periods and magnitudes compared 
to the rest of our data.  If V01 in And~I is a P2C, then it will mean that 
we have been unable to find a definite AC in this dSph galaxy.  And~I might 
then become the first dSph galaxy without an AC, 
though we have not by any means surveyed all of the galaxy for variable 
stars (see Table~9).  We also note, as was first shown by DAC02 using 
a different ``shorter" period,  that it is possible to fit the 
observations for And~III V08 with a period as short as $P=0.655$~d.  
This would make the star an AC, although this is not the best $\chi^2$ 
fit according to the template fitting program we have employed.  Given 
the lack of phase coverage, we do not feel comfortable forcing a change in the 
period soley to make it an AC.  The mean magnitudes we find for this short 
period would place the variable on the fundamental mode period-luminosity 
relation line for the ACs.  We were unable to find a viable ``AC-like'' 
period for And~I V01.

Do we expect to see P2C variables in dSph galaxies?  In Galactic globular 
clusters, the main property associated with the presence of P2Cs are 
strong blue HBs (e.g., DAC02; Wallerstein 2002).  In particular, the 
shorter period P2Cs, known as BL~Her stars, which And~I V01 and And~III V08 
would be classified as, are assumed to be post blue HB stars evolving 
through the instability strip towards the asymptotic giant branch.  Of 
the Galactic dSph galaxies, only Fornax has been shown to contain P2Cs 
(Bersier \& Wood 2002).  It may be that Fornax has a significant population 
of blue HB stars that could provide a source for P2Cs, but it is difficult 
to infer this from the CMD as the blue HB region is severely contaminated 
by main sequence stars.  On the other hand, the CMDs for And~I and 
And~III clearly do not contain strong blue HB populations (see Figure~1) 
and this would seem to argue against the P2C interpretation.  We are 
forced to conclude that further observations of these stars are 
needed to clarify their periods and light curve shapes, and thus their 
classification.

\begin{figure*}[t]
  \centerline{\psfig{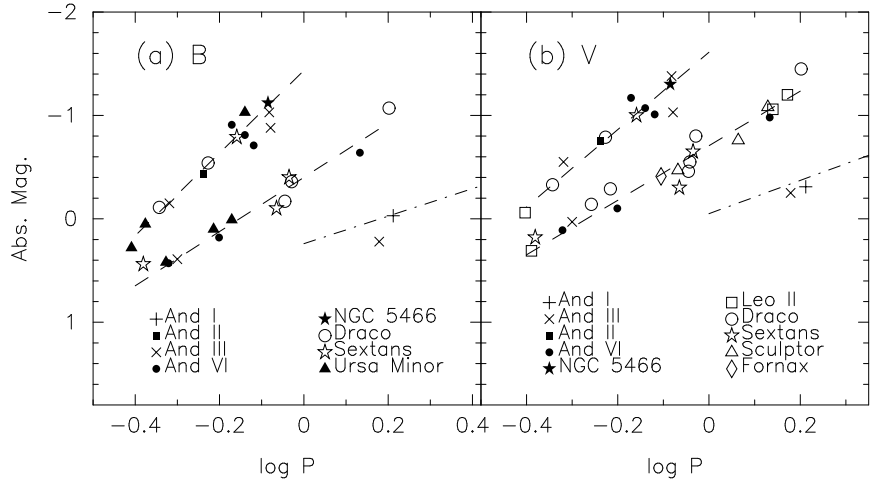}}
  \caption{Period-luminosity diagrams for anomalous Cepheids for both the 
   (a) $B$ and (b) $V$ filters.  The anomalous Cepheid fundamental
   and first-overtone mode period-luminosity relations from Pritzl et al.\ (2002)
   are shown as dashed lines.  The Population~II Cepheid period-luminosity line
   from McNamara (1995) is shown as a dotted-dashed line.}
  \label{Fig06}
\end{figure*}

\subsection{The Frequency of Anomalous Cepheids in dSph Galaxies} 

Mateo et al.\ (1995) pointed out that the specific frequency of
ACs (the number of ACs per $10^5 L_{V,\odot}$) correlates with both 
(visual) luminosity and mean abundance for the Galactic dSphs.  In Paper~II
we updated the Galactic dSph data and found that the M31 dSphs And~II and
And~VI apparently follow the same correlations.  We now explore whether
this result applies also to And~I and And~III\@.  In Table~9 we list the data 
used in calculating the specific frequency of ACs for  
And~I and And~III (cf.\ Table~5 of Paper~II).  We note that in this 
discussion, the supra-HB variable stars in And~I and And~III that 
may be ACs or P2Cs, have been assumed to be ACs.  Without this assumption, 
And~I would have no ACs and And~III would have a specific frequency of 1.7 
per $10^5 L_{V,\odot}$.  Figure~7 plots the logarithm of specific frequency
against M$_{V}$ and mean metallicity for And~I and And~III along with the 
data from Paper~II\@.  We also include the 
specific frequency value for the Phoenix dwarf irregular galaxy from 
Gallart et al.\ (2004).  Clearly, like And~II and And~VI (cf. Paper~II),
And~I and III fall near the line defined by 
the Galactic dSph galaxies, as does Phoenix (cf.\ Gallart et al.\ 2004).  
Consequently, 
it appears that the correlation between AC specific 
frequency and either (visual) luminosity or metallicity seen in the 
Galactic dSph galaxies applies also to the M31 systems. 

As pointed out by Mateo et al.\ (1995) and in Paper~II, it is perhaps 
surprising to see a strong correlation between the specific frequency of 
ACs and the total luminosity of the dSph galaxies given the stellar 
population differences among the dSphs, and the fact that there are two 
different mechanisms that can create ACs, one 
from younger stars and the other from mass-transfer in binaries.  Both 
mechanisms are complicated and both can occur in a dSph galaxy, as 
discussed by Bono et al.\ (1997) using various stellar models.  However, 
as noted by Mateo et al.\ (1995) and in Paper~II, the mean abundances 
of dSph galaxies are well correlated with their luminosities, with the 
lower luminosity systems having lower abundances.  Thus, since regardless 
of the mechanism creating the AC the star needs to be metal-poor to 
enter the instability strip, it may simply be that it is easier to 
generate ACs in a more metal-poor environment.

We now comment briefly on the observed lack of ACs in the M31 halo field 
surveyed for variable stars by Brown et al.\ (2004).    
Given the results of Durrell, Harris, \& Pritchet (2004), for example, we
can assume that the mean metallicity for the surveyed region is
approximately $\langle {\rm [Fe/H]} \rangle = -0.8$.  Then assuming that 
the relation between AC specific
frequency and mean abundance shown in Figure~7 can be extended to mean
abundances of this order, the linear fit shown gives $\log\,S\sim-1.55$ 
per $10^5 L_V$.  The mean surface brightness and area of the region
(Brown et al.\ 2004), then yield, assuming $(m-M)_V=24.5\pm0.1$~mag,  
a visual luminosity for the region of $7.4 \times 10^5 L_V$.  Therefore, 
the number of ACs expected in the field is $\sim0.2$.  This is certainly 
consistent with the fact that Brown et al.\ (2004) found none.  However, 
we know that ACs derive from metal-poor stellar populations 
(${\rm [Fe/H]}<-1.3$; Demarque \& Hirshfeld 1975).  Consequently, using 
a mean metallicity of ${\rm [Fe/H]}=-0.8$ may not be the most appropriate 
approach.  
Using Eq.~2 of Alcock et al.\ (2000), Brown et al.\ (2004) found that the RRL 
population in their field has a mean metallicity of $-1.79$.  For this 
metallicity, Figure~7 gives $\log\,S\sim-0.1$ per $10^5 L_V$.  However, in this
situation, we cannot apply the total visual luminosity for the surveyed 
region.  Brown et al.\ (2003) argue that, in this region, the old, metal-poor 
stellar population makes up at most 30\% of the total 
(cf.\ Durrell et al.\ 2004).  Scaling the total luminosity by this
fraction then predicts 1.8 ACs in the surveyed field.  This 
number is not inconsistent with the observed lack of ACs 
given small number statistics and the uncertainty in the luminosity 
fraction generated by the metal-poor population.  Clearly further variable 
star
searches over larger areas are required to more fully investigate 
whether or not ACs exist in the M31 halo.  A confirmed lack of 
such stars would add  strength to the contention that the disruption
of dwarf galaxies 
comparable to the existing M31 dSph companions was not a major contributor 
to the build-up of the M31 halo.

\begin{figure*}[t]
  \centerline{\psfig{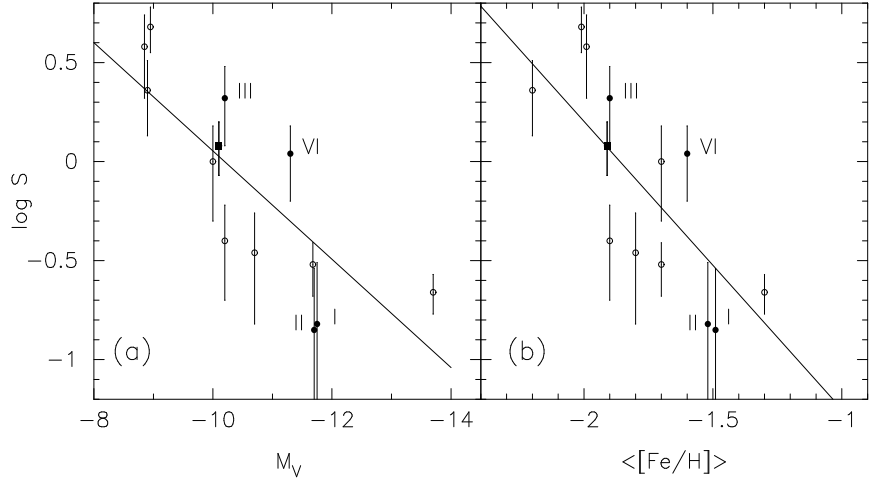}}
  \caption{Specific frequency of anomalous Cepheids in dwarf spheroidal galaxies 
   as a function of the absolute visual magnitude (left panel),
   and the mean metallicity of the dwarf galaxy (right panel).  The Galactic dwarf
   spheroidal galaxies are shown as open circles, while the M31 ones are shown as
   filled circles.  The Phoenix dwarf galaxy is shown as a filled square.  In both plots,
   there are clear trends that the Galactic dwarf spheroidal galaxies follow
   as shown by the least squares fit to the data.  The M31 dSph
   galaxies follow these trends, including And~I and And~III\@.  Note that we have
   assumed that the two supra-HB variables stars And~I V01 and And~III V08 to
   be anomalous Cepheids, although their classification is uncertain.}
  \label{Fig07}
\end{figure*}

\subsection{Anomalous Cepheids $=$ Short-Period Cepheids?} 

Recently, the number of variable star surveys of dwarf galaxies has 
increased, especially for dwarf irregular galaxies such as IC~1613 
(Dolphin et al.\ 2001), Leo~A (Dolphin et al.\ 2002), Sextans~A 
(Dolphin et al.\ 2003), NGC~6822 (Clementini et al.\ 2003) and 
Phoenix (Gallart et al.\ 2004).  These studies have revealed a number of short-period 
(Classical) Cepheids.  Given that these variables 
have periods similar to the ACs and comparable magnitudes, 
there is some question as to the distinction between the two classes of 
variables.  Using the OGLE Small Magellanic Cloud database (Udalski et al.\ 1999), 
Dolphin et al.\ (2002) found the P-L relations for Cepheids with periods 
of two days or less was similar to the AC P-L relations of Nemec, Nemec, 
\& Lutz (1994).  Therefore, they concluded that there was no clear distinction between 
ACs and short-period Cepheids in terms of their position 
in the P-L relations.  The only difference between these two 
classes would be that short-period Cepheids are predominantly found in 
systems with young stars (ages$<100$~Myr) while ACs preferentially 
occur in systems that lack such young stars. 

Baldacci et al.\ (2003) were the first to show that the short-period 
Cepheids in dwarf irregular galaxies, including the Large and Small 
Magellanic Clouds, have P-L relations different than those of ACs as 
given in Paper~I\@.  They also comment that Phoenix and NGC~6822 may have their 
instability strip continuously populated from the RRL stars, through the ACs 
to the short-period Cepheids.  Marconi, Fiorentino, \& Caputo (2004) and Gallart et al.\ 
(2004) took this a step further claiming that some of the short-period 
Cepheids in dwarf galaxies such as Leo~A and Sextans~A may actually be 
ACs.

Following the same method as outlined in \S4 of Paper~I, we determined the absolute 
magnitudes of the short-period Cepheids in IC~1613, Leo~A, and 
Sextans~A using the tip of the red giant branch as the distance 
indicator.  This is to ensure consistency between the absolute magnitudes 
found for the ACs and the short-period Cepheids.  We 
plot the absolute magnitude of the short-period Cepheids against their 
periods in Figure~8 along with the lines for 
the AC P-L relations as determined in Paper~I (cf.\ Eqns.~4 and 5).  Looking 
first at the fundamental mode variables, which are found at a fainter 
magnitude than the first-overtone variables for a given period, the 
short-period Cepheids are clearly found at higher luminosity than the 
ACs at a given period.  For 
the first-overtone Cepheids, the case isn't as clear.  However for a set 
period, these stars are still offset toward brighter absolute magnitudes 
than the first-overtone ACs, especially at short periods.  There are some 
stars which are located near the AC P-L lines, but this may be due to the 
uncertainties in the periods and magnitudes of the stars.  The higher 
luminosity at fixed period for the short-period Cepheids relative to 
the ACs indicates that these stars have higher masses than the ACs.  This 
is not unexpected given that these dwarf irregular galaxies have on-going 
star formation which the dSph galaxies lack.

Our analysis agrees with Baldacci et al.\ (2003), Marconi et al.\ (2004), 
and Gallart et al.\ (2004) as regards to the existence of a difference 
between the P-L relations for ACs and short-period Cepheids.  In essence, 
the reason we do not see as many short-period Cepheids lying along the AC 
P-L relations as do Dolphin et al.\ (2001, 2002, 2003) lies with their 
use of the Nemec, Nemec, \& Lutz (1994) AC P-L relations rather than 
those of Paper~I.\@  This does not mean, however, that dwarf galaxies such as 
Leo~A and Sextans~A do not contain both short-period and anomalous 
Cepheids.  Indeed, we encourage further observations of these and 
other dwarf irregular galaxies to search for such stars.

\begin{figure*}[t]
 \centerline{\psfig{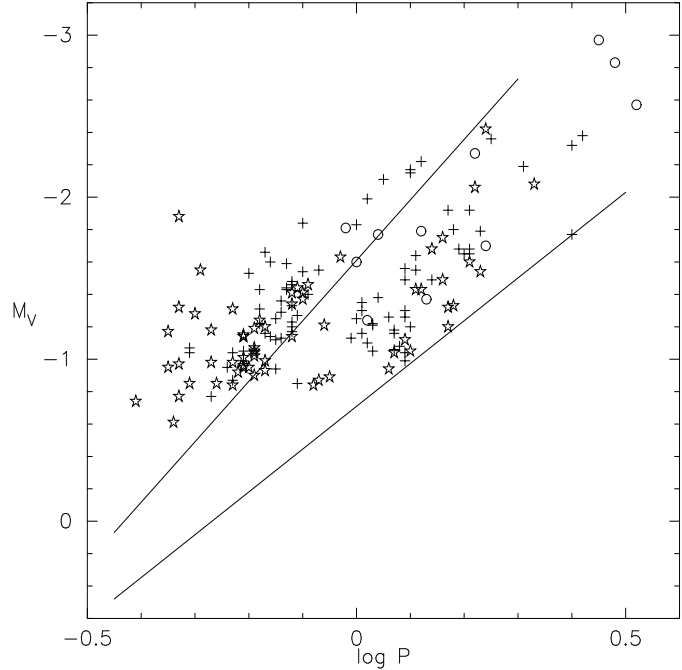}}
 \caption{$V$-band absolute magnitude versus the logarithm of the period 
  for short period Cepheids in Sextans~A (plusses),
  IC~1613 (open circles), and Leo~A (open stars).  The solid lines are
  the period-luminosity relations representing the fundamental and
  first-overtone mode anomalous Cepheids (Pritzl et al.\ 2002).  This plot
  clearly illustrates that the short period Cepheids are found at higher
  luminosities than the anomalous Cepheids at a given period.}
 \label{Fig08}
\end{figure*}

\section{RR Lyrae Stars} 

\subsection{And~I} 

We detected 99 RRL stars in our field-of-view of And~I, with 72 pulsating in 
the fundamental mode (RRab) and 26 pulsating in the first-overtone mode (RRc).  
This nearly doubles the number of RRL stars found by Da~Costa et al.\ (1997), 
who detected 50 RRL stars.  One variable, V04, remains unclassified due to a lack 
of F555W data because of a chip defect.  It is likely an RRL star given its 
F450W magnitude.  The presence of the RRL stars indicates that 
And~I contains a stellar population with an age $> 10$~Gyr.  It should be noted 
that due to the low amplitudes of the RRc stars, we may not have detected all 
of the possible stars pulsating in this mode.   

We find the mean period of the And~I RRab and RRc stars to be 0.575~day and 
0.388~day, respectively.  The ratio of RRc stars to the total number of 
RRLs, $N_c/N_{RR}$, is 0.27.  If we place these values in Table~6 of Paper~I 
according 
to the metallicity given in DACS00, $\langle {\rm [Fe/H]} \rangle = -1.46 \pm 
0.12$, we find that they are consistent with those found for other 
dSph galaxies.  With this metallicity, the RRL stars in And~I also follow 
the mean period - metallicity relation defined by the Galactic globular 
clusters, 
as is shown in Figure~9.  The mean period for the RRab stars in And~I is close 
to the value found for And~II which has a very similar mean metallicity 
to And~I (DACS00).  It is interesting to note that this is the case even 
though there is a large spread among the And~II RRab stars in the 
period-amplitude diagram (see Fig.~7 in Paper~II).

\begin{figure*}[t]
 \centerline{\psfig{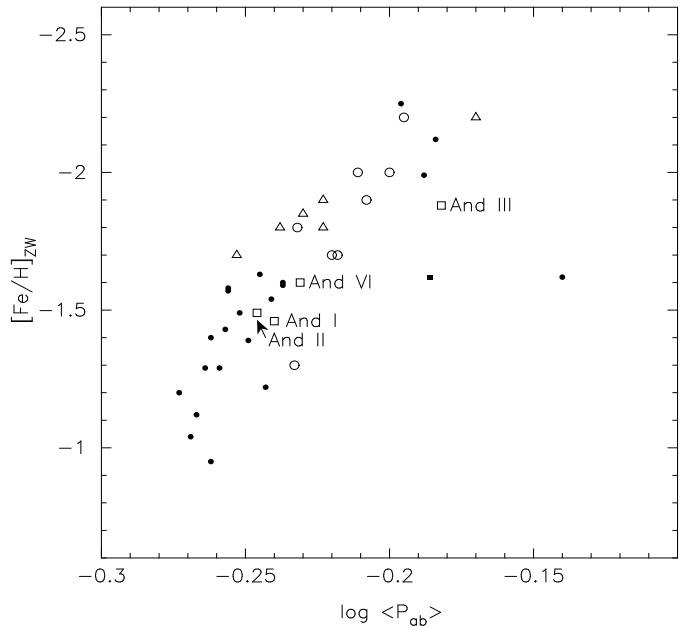}}
 \caption{Mean period for the RRab stars versus the mean metallicity of the 
  parent system.  The Andromeda dwarf spheroidal galaxies
  are shown as open squares.  The Galactic dwarf spheroidals are indicated
  by open circles.  Galactic globular clusters with at least 15 RRab stars
  are shown as filled circles, along with $\omega$~Centauri (filled square).
  Large Magellanic Cloud globular clusters with a minimum of 15 RRab stars
  are shown as open triangles.}
 \label{Fig09}
\end{figure*}

The mean magnitude of the And~I RRL stars, not including the two fainter 
RRL stars, is found to be $\langle V_{\rm RR} \rangle = 25.14\pm0.04$~mag, 
where the uncertainty is the aperture correction uncertainty, the photometry 
zeropoint uncertainty in the DAC02 photometry, and the uncertainty in the 
mean RRL instrumental magnitude derived from the spline-fitting routine added 
in quadrature to the standard error of the mean.  This matches the value 
determined by DACS00 of $V_{\rm HB} = 25.23\pm0.04$~mag.  Given $\langle {\rm [Fe/H]} 
\rangle = -1.46 \pm 0.12$, the equation $M_{V,{\rm RR}} = 0.17{\rm [Fe/H]} + 0.82$ 
from Lee, Demarque, \& Zinn (1990) yields $M_{V,{\rm RR}} = +0.57$ and thus a 
true distance modulus of $24.42 \pm 0.07$~mag for the adopted reddening.  
This corresponds to a distance of $765\pm25$~kpc in good accord with the distance, 
$790\pm25$~kpc, found by DACS00.

\subsubsection{M31 Halo RR~Lyrae Stars?} 

\begin{figure*}[t]
  \centerline{\psfig{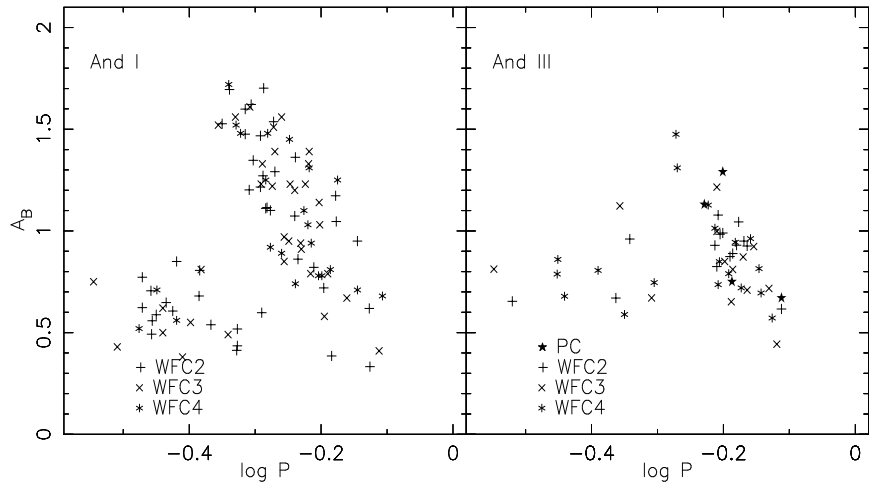}}
  \caption{Period-amplitude diagram for the RR~Lyrae stars in And~I (left panel)
   and And~III (right panel) for the $B$ filter.}
  \label{Fig10}
\end{figure*}

According to DACS96 and DACS00, the line-of-sight distance between M31 and 
And~I is $0\pm70$~kpc and the true distance of And~I from the center of M31 
is somewhere between $\sim 45$ and $\sim 85$~kpc.  This makes And~I 
the closest dSph to M31 leading to the possibility that there may 
be contamination of the CMD from M31 halo stars.  In fact,  DACS96 showed that a 
number of stars somewhat fainter and redder than the And~I red giant branch
(RGB) in their 
CMD were likely to be from the M31 halo. 

Variable stars V76 and V100 in And~I are both clearly fainter than the rest 
of the RRL in this dSph galaxy.  We checked the images for any defects 
near these stars which may have affected their photometry and none 
were found.  DACS96 drew attention to a population of faint blue stars 
in And~I which they suggested might be blue straggler stars analogous to 
those found in globular clusters.  Some blue straggler stars are known to 
be short period variables ($\delta$~Scuti), but the periods of the And~I 
faint variables are too long for them to be interpreted in this way.  
Therefore, it appears that V76 and V100 must originate from a stellar 
population not associated with And~I\@.  The only logical interpretation 
of this is that they are part of the M31 halo.

We can attempt to verify this assertion by making use of the recent study 
of RRL stars in the halo of M31 of Brown et al.\ (2004).  These authors 
used the Advanced Camera for Surveys on HST to provide a complete census of 
RRL stars in a halo 
field approximately 51 arcmin from the center of M31 on the minor axis.  
In their field the density of RRab variables is 2.24 per ${\rm arcmin}^2$.  
Brown et al.\ (2004) also give the surface brightness of the M31 halo at
the location of their field as 26.3 $V$ mag per ${\rm arcsec}^2$.  Using 
the Pritchet \& van den Bergh (1994) surface brightness profile, DACS96 
estimated 
the M31 halo surface brightness in the vicinity of And~I as approximately 
29.5 $V$ mag per ${\rm arcsec}^2$.  Then assuming no change in the stellar 
population 
with radius, these figures imply a M31 halo RRab density at the location of 
And~I 
of approximately 0.12 per ${\rm arcmin}^2$.  Combined with the area of the 
field studied 
here, this density predicts only 0.6 M31 halo RRab should be present in our 
data, rather than the two observed.  Consequently, even allowing for small 
number statistics, it appears that these two fainter RRL may be associated 
with a region of moderately enhanced density in the halo of M31.  This conclusion is 
strengthened by looking at the magnitude distribution of the RRab stars in the 
Brown et al.\ (2004) M31 halo field.  The 29 stars have a mean $V$ magnitude, 
corrected to the reddening of the And~I, of $\langle V \rangle=25.18$ mag, 
with a standard deviation of 0.11 mag (in F606W, which for this purpose we 
can equate with $\sigma_V$).  The largest excursion from this mean on the 
faint side is only 0.24 mag.  Consequently, the two fainter RRab stars in the 
And~I field are both more than 3$\sigma$ fainter than the mean, increasing the 
likelyhood that they are associated with a distinct feature, rather than the 
general field in the outer halo of M31.

This indeed may be the case.  McConnachie et al.\ (2003) have used the CHF12k 
CCD mosaic camera to map the giant stellar stream in the halo of M31 
(Ibata et al.\ 2001; Ferguson et al.\ 2002) to larger radial distances.  They 
find the stream extends to the South-East of M31 for at least 4.5$^\circ$, 
and using the tip of the red giant branch as a distance indicator, 
McConnachie et al.\ demonstrate that the Stream is on the far side of M31, 
increasingly so as the angular distance from M31 increases.  Coincidently, 
And~I is 
projected close to this Stream on the sky, and, based on Fig.\ 7 of 
McConnachie et al., we estimate that the Stream lies approximately 150 kpc 
beyond M31 (and And~I, which is at the same distance as M31 [DAC02]) at the 
location of And~I\@.  This corresponds to a total distance of $\sim$875 kpc 
or an apparent visual distance modulus of approximately 24.9~mag.  If the two 
faint RRL stars are members of this Stream, then their absolute visual 
magnitudes would be $\sim$0.7 (V76) and $\sim$0.9 (V100).  Such values are 
not unexpected given the apparent metal-rich nature of the Stream 
(Ibata et al.\ 2001; Ferguson et al.\ 2002; McConnachie et al.\ 2003).  
We conclude then that it is quite possible that the two faint RRL stars 
observed in the field of And~I are associated with the recently identified 
Stream in the halo of M31, rather than with the general M31 halo field.
The existence of RRL stars in the Stream immediately indicates that its
progenitor must have contained an old stellar population, and such variables 
may provide a further means to map both the location of the Stream and to
investigate its stellar population.

\subsection{And~III} 

For And~III, we found a total of 51 RRL, 39 pulsating in the fundamental mode 
and 12 in the first-overtone mode.  As 
mentioned above, it is more difficult to detect the RRc stars due to their 
smaller amplitudes.  Therefore not all of the RRc stars within our 
field-of-view may have been detected.  

The mean periods are 0.657~day and 0.402~day for the RRab and RRc stars, 
respectively.  The $N_c/N_{RR}$ ratio is found to be 0.24, although 
this number and the mean period for the RRc stars may be uncertain due to 
undetected RRc stars.  As was done above for And~I, we can place these values
in Table~6 of Paper~I according to And~III's metallicity of $\langle {\rm [Fe/H]} 
\rangle = -1.88\pm0.11$.  We find that in comparison to the other dSph 
galaxies, And~III has unusually long periods for its RRab and RRc stars.  
This can clearly be seen in the mean period - metallicity diagram of Figure~9 
where And~III lies to the right of 
the other dSph galaxies and globular clusters with similar abundances.  
A possible reason for this will be discussed in \S5.3. 

The mean magnitude for the And~III RRL stars is $\langle V_{\rm RR} \rangle = 
25.01\pm0.04$~mag.  DAC02 found the mean $V$ magnitude of the HB of And~III 
to be $25.06\pm0.04$~mag.  Given $\langle {\rm [Fe/H]} \rangle = 
-1.88\pm0.11$, $E(\bv)=0.06\pm0.01$, and the equations used above, we find 
$M_{V,{\rm RR}} = +0.50$ and $A_V = 0.17\pm0.03$.  This yields a distance 
of $740\pm20$~kpc, which matches quite well with the estimate of 
$750\pm20$~kpc by DAC02.

\subsection{RR~Lyrae Period-Amplitude Diagrams}

Figure~10 plots the $B$-amplitude of the RRL stars against their adopted
period for And~I and And~III\@.  As discussed in Paper~I, the scatter in these 
diagrams was reduced by revising the period of a small number of stars where
it was apparent that A. C. Layden's period-finding routines had picked out an 
alias of the probable true period.  For And~I the longer duration of the 
observation set meant that there were only minimal period revisions for this 
dSph.
Figure~10a also shows that in And~I there are a number of longer period RRc 
stars which overlap with similar period, but higher amplitude, RRab stars.   

In Figure~10b, we see that the RRab stars in And~III have mostly smaller 
amplitudes and longer periods.  In direct contrast to And~I, there is a
notable lack of shorter period, higher amplitude RRab 
stars.  It is this lack of shorter period RRab stars that results in 
a relatively long mean period, thus displacing the And~III point in the
mean period - metallicity diagram (Figure~9) to the right.  

The explanation may lie with the morphology of the horizontal branch 
in this dSph.  It is clear from the CMD of And~III (see Fig.~1b; Fig.~3 in 
DAC02) that the horizontal branch is lacking in stars bluer than the 
instability strip.  A majority of the stars are located toward the 
red side of the horizontal branch and therefore, more of the red side 
of the instability strip is filled.  As a consequence, longer period 
RRab and RRc stars are likely.  A similar situation is 
thought to occur for the RRL variables in the globular cluster Ruprecht~106 
(Catelan 2004), a cluster which also has a very red HB.

\begin{figure*}[t]
 \centerline{\psfig{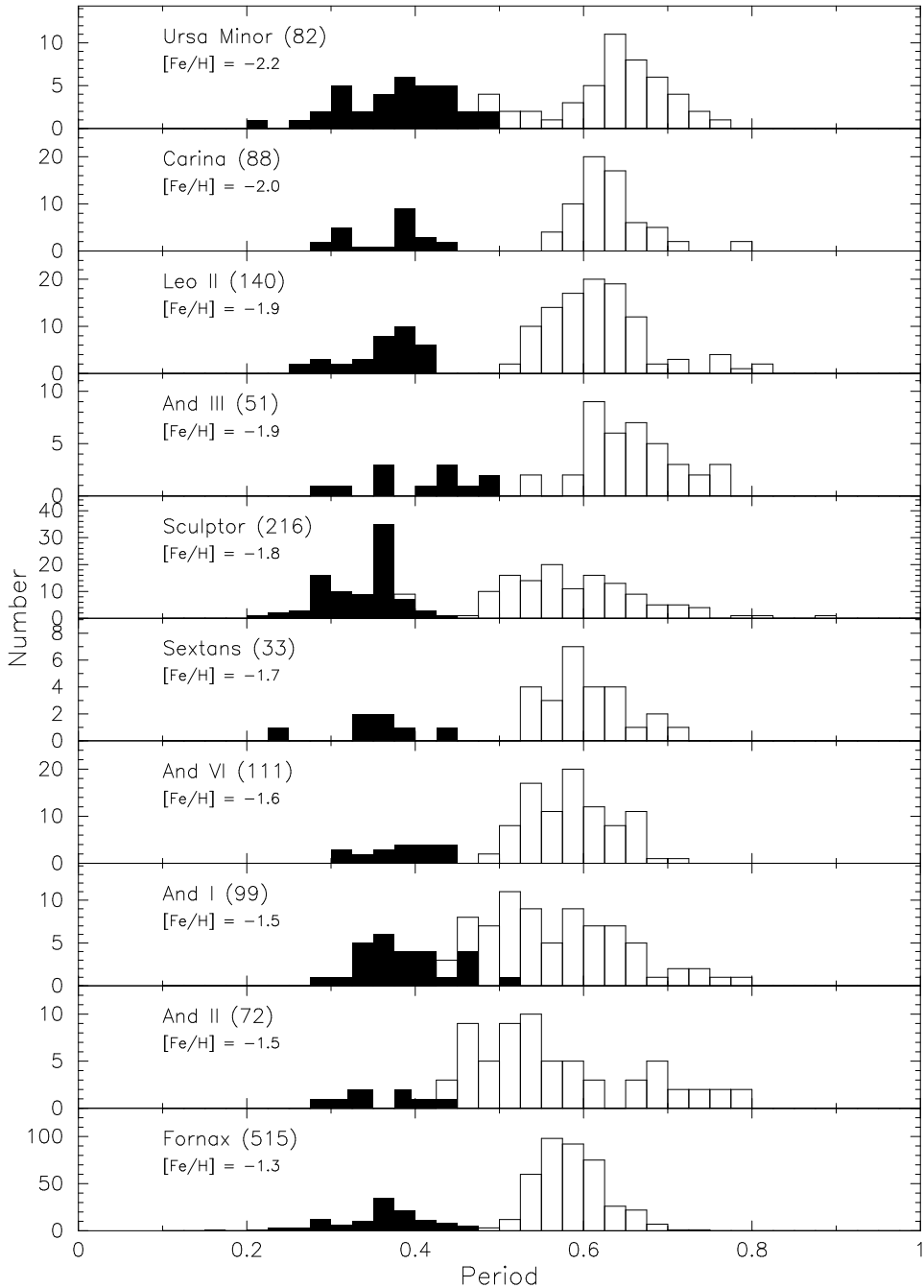}}
 \caption{Period distribution plots for the RR~Lyrae stars in dwarf spheroidal
  galaxies.  The plots are arranged so that the mean metallicity
  of the parent system increases from the top down.  RRab stars are shown as
  open histograms while RRc stars are shown as filled histograms.}
 \label{Fig11}
\end{figure*}

\begin{figure*}[t]
 \centerline{\psfig{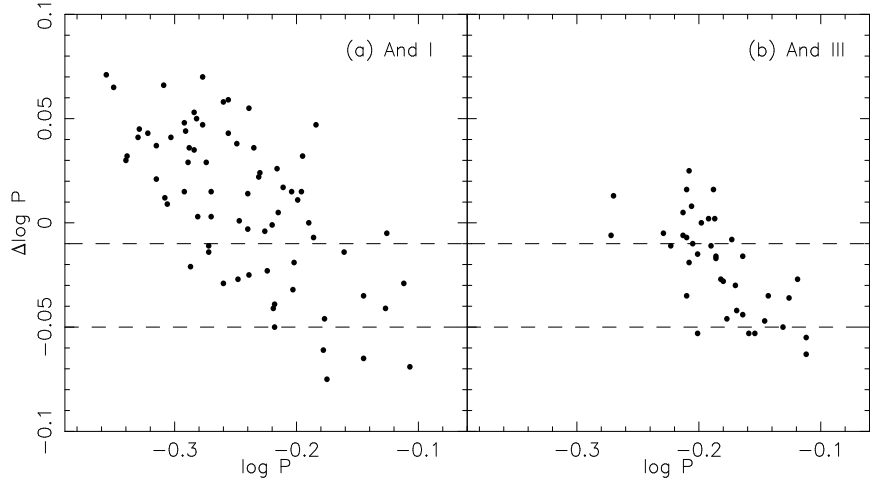}}
 \caption{Period shift versus period for the RRab stars in (a) And~I and
  (b) And~III\@.  The dashed lines represent the zone in which few
  RRab stars from Galactic globular clusters are found.}
 \label{Fig12}
\end{figure*}

\begin{figure*}[t]
 \centerline{\psfig{figure=Pritzl.fig13.ps,height=5.25in,width=3.50in}}
 \caption{Upper panel:  [Fe/H] distribution plot for the RRab stars in
  And~I\@.  The individual [Fe/H] values were calculated using Eq.\ 2 and have
  an uncertainty of $\sim0.3$~dex.  The dashed line is a gaussian fit to these data.
  It has a mean of ${\rm [Fe/H]}=-1.44$ and a standard deviation of
  0.51~dex.  Lower panel:  [Fe/H] distribution plot for red giant branch stars in And~I
  from DACS00.  The individual [Fe/H] values have uncertainties of $\sim0.1$~dex.}
 \label{Fig13}
\end{figure*}

\begin{figure*}[t]
 \centerline{\psfig{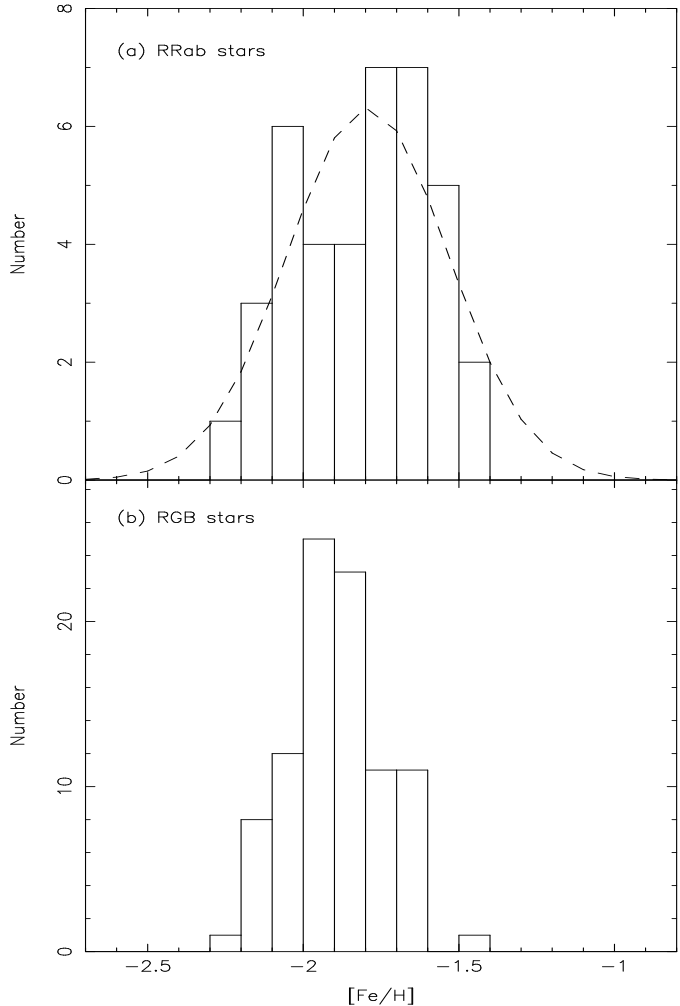}}
 \caption{Upper panel:  [Fe/H] distribution plot for the RRab stars in And~III\@.
  The individual [Fe/H] values were calculated using Eq.\ 2 and have
  an uncertainty of $\sim0.3$~dex.  The dashed line is a gaussian fit to these
  data.  The curve has a mean of ${\rm [Fe/H]}=-1.79$ and a standard
  deviation of 0.37~dex.  Lower panel:  [Fe/H] distribution plot for red giant branch
  stars in And~III from DAC02.  The individual [Fe/H] values have uncertainties of $\sim0.1$~dex.}
 \label{Fig14}
\end{figure*}

\subsection{RR~Lyrae Period Shift} 

Another way to compare the RRL populations is to view their period 
distribution.  In Figure~11 we plot the RRL period distribution for a number 
of dSph galaxies according to their mean metallicity.  As with the mean 
RRab periods discussed above, there is a trend for the RRab distribution to
shift toward shorter periods as the metallicity increases.  For And~III, 
the distribution has a marked discontinuity in the RRL numbers at $P \sim 0.6$~day, 
a further manifestation of the lack of shorter period, higher amplitude RRab stars 
in this dSph galaxy.  

The shift in period among the RRab stars in globular clusters was first 
noticed by Sandage, Katem, \& Sandage (1981) and was later found to be 
dependent on metallicity by Sandage (1981; 1982a,b).  With M3 as the fiducial 
cluster, the period shifts of RRab stars were determined to follow the 
relation: 
\begin{equation} 
\Delta\log\,P = -(0.129A_B + 0.088 + \log\,P)
\end{equation}
\noindent 
by Sandage (1982a,b).  Using this equation, Sandage found that more metal-rich 
globular clusters have $\langle\Delta\log\,P\rangle \ge -0.01$, while the more 
metal-poor clusters have $\langle\Delta\log\,P\rangle \le -0.05$ with few 
globular cluster RRab stars found between these values.  However, a number 
of RRab stars in dSph galaxies have been found 
in this gap (Sextans:  Mateo, Fischer, \& Krzeminski 1995; Leo~II:  Siegel \& Majewski 
2000; Andromeda~VI:  Paper~I; Andromeda~II:  Paper~II).  The presence of 
these stars in the gap indicates that the RRL populations in dSph galaxies 
are not simply superpositions of two separate populations, one being 
metal-rich and the other metal-poor.  In column~10 of Tables~2 and 3, 
we list the $\Delta\log\,P$ values for the RRab stars of And~I and 
And~III\@.  These values are plotted against the corresponding 
$\log\,P$ values in Figure~12.  Clearly, like the other dSph galaxies, both 
And~I and And~III have a substantial number of RRab stars within the gap.  
This figure also shows the lack of short period RRab stars in And~III\@.  
The mean period shift is +0.01 for And~I and --0.02 for And~III\@.  Using 
$\Delta\log\,P = 0.116 {\rm [Fe/H]} + 0.173$ from Sandage (1982a), we 
find ${\rm [Fe/H]} = -1.41$ for And~I and $-1.66$ for And~III\@.  The 
value determined for And~I is close to that found by DACS00 
($\langle{\rm [Fe/H]}\rangle = -1.46\pm0.12$) based on the colors of RGB 
stars.  On the other hand, the 
estimate for And~III is somewhat more metal-rich than that given by DAC02 from 
the color of the RGB ($\langle{\rm [Fe/H]}\rangle = -1.88\pm0.11$).  This 
discrepancy is reduced with the metallicity estimates made in the following 
section using different methods.

\subsection{Metallicity Estimates from the RR~Lyrae Stars}

We found in Papers~I and II that the equation: 
\begin{equation} 
{\rm [Fe/H]}_{\rm ZW} = -8.85(\log\,P_{ab} + 0.15A_V) - 2.60
\end{equation}
\noindent 
derived by Alcock et al.\ (2000) using the RRab stars in M3, M5, and M15 
yields a mean metallicity for the RRL in And~II and And~VI similar 
to those found using the mean colors of the RGB\@.  The accuracy of
this equation is $\sigma_{\rm [Fe/H]} = 0.31$ per star (Alcock et al. 2000).  
Using 
this equation, we derived individual metallicities for the RRab stars and
these are
listed in column~11 in Tables~2 and 3 for And~I and And~III, respectively.

The upper panel of Figure~13 plots the abundance distribution for the RRab stars 
in And~I using the individual abundances derived from Eq.~2.  The mean and 
median of the values are ${\rm [Fe/H]} = -1.49$ and $-1.46$, respectively.  
A gaussian fit to these data is also shown which has a mean of 
${\rm [Fe/H]} = -1.44\pm0.03$ (standard error of the mean).  These RRab 
based abundances agree well with the mean And~I abundance, ${\rm [Fe/H]}=-1.46\pm0.12$, 
determined by DACS00 from the colors of the red giant branch stars.  As noted 
in Paper~II, because not every red giant branch star evolves to become 
an RRab, there is no obvious reason to expect these mean abundances to agree.

The gaussian fit shown in the upper panel of Fig.~13 has a standard deviation of 0.51~dex.  
Given that the individual RRL abundances have uncertainties of order 0.3~dex, 
this suggests the presence of a real abundance dispersion among the And~I 
RRab stars, with $\sigma_{{\rm [Fe/H], int}}\sim0.4$~dex.  This value is somewhat 
larger than the abundance dispersion found by DACS00 from the And~I red giant 
branch ($\sigma_{{\rm [Fe/H], int}}\sim0.21$) though, again recognizing that not 
every red giant branch star evolves to become a RRab star, it is difficult 
to compare these dispersions in a meaningful way.  The lower panel of Fig.~13 
shows the red giant branch abundance distribution for And~I from DACS00.  
Given that the individual red giant branch abundance determinations have substantially 
lower uncertainties than the RRab values, the distributions appear quite 
comparable.  Both distributions appear to be sharply limited 
on the metal-rich side.  The And~I results are similar to the situation for 
And~II, where both the RRab and the red giant branch stars show substantial 
abundance ranges, larger indeed than those found here for And~I (Paper~II).

For And~III, the distribution of the individual metallicity estimates for the 
RRab stars is shown in the upper panel of Figure~14.  Here the mean and 
median metallicity values are $-1.81$ and $-1.77$, respectively.  The gaussian fit to the 
data has a mean value of $-1.79\pm0.03$ (standard error of the mean).  
Once again, despite the fact that only a relatively small fraction of the red 
giant branch stars evolve to become RRab stars (cf.\, Fig.~1), these RRab 
based abundance values agree well with the red giant branch based determination 
of the mean abundance, ${\rm [Fe/H]}=-1.88\pm0.11$, given by DAC02.

The gaussian fit shown in Fig.~14 has a standard deviation of 0.37~dex, 
considerably smaller than was found for And~I, and barely larger than that 
expected on the basis of the individual errors alone ($\sim0.3$~dex).  
There is therefore no compelling case for the existence of a substantial 
abundance spread among the And~III RRab stars.  This result is similar 
to that found for the RRab in And~VI (Paper~I).  Again while there is no 
reason to expect this to be the case, the small abundance dispersion for the 
And~III RRab stars is consistent with the results for the red giant branch:  
DAC02 give $\sigma_{{\rm [Fe/H], int}}=0.12$ from their analysis of the 
red giant branch colors.  The lower panel of Fig.~14 reinforces this conclusion:  
it shows the red giant branch abundance distribution from DAC02.  As for 
Fig.~13, given the larger uncertainty in the individual RRab values, the 
distribution in the upper and lower panels appear quite similar.

Now that we have used the Alcock et al.\ (2000) equation (Eq.~2) to estimate 
the mean metallicity for the RRL in four of the M31 dSph galaxies, it is 
worthwhile to evaluate the outcome of this process.  Even though the 
metallicity estimate for any individual RRL from this equation is of 
limited accuracy, for each of the four dSph galaxies the mean metallicity 
derived from the RRL stars is in excellent accord with that determined 
from the RGB\@.  Indeed, for these four systems, the standard deviation of 
the differences between the RRL and RGB metallicity estimates is only 
0.05~dex.  Since not every RGB star evolves into an RRL, there is no 
{\it a priori} reason to expect this good agreement.  Yet clearly the RRL stars, 
 and the use of Eq.~2, are providing valid estimates of the mean 
metallicities of these four systems.  Consequently, application of Eq.~2 
may prove worthwhile in those situations where information is available 
only for RRL stars.  Further applications to other dSph galaxies and globular 
clusters will be important to test the application of this equation.

\section{Summary \& Conclusions} 

We have presented the results of variable star surveys of the M31 dSph 
galaxies And~I and And~III using the HST/WFPC2.  And~I was found 
to contain 72 fundamental mode and 26 first-overtone mode RRL stars 
with mean periods of 0.575~day and 0.388~day, respectively, with one 
unclassified RRL star.  We also found that two of the RRL in the 
And~I field are fainter than the HB; these stars 
may be associated with the recently discovered stellar stream in the halo 
of M31.  In And~III, we discovered 39 fundamental mode and 12 
first-overtone mode RRL stars with mean periods of 0.657~day and 0.402~day, 
respectively.  The mean period of the RRab stars in And~I is consistent 
with the Galactic globular cluster RRL mean period - metallicity relation.  For 
And~III, the mean period of the RRab stars is longer than that expected 
for the mean metallicity of the galaxy.  This discrepancy arises from a 
lack of shorter period, higher amplitude RRab stars in And~III 
compared to other dSphs.  This effect may be related to the very red 
morphology of the HB in this dSph.  
We also find for both dSph galaxies that the mean magnitudes of the RRL 
imply distances similar to those previously determined from the mean 
magnitude of the horizontal branch.

We found five variable stars that are brighter than the HB in And~III, four 
of which are ACs.  These stars have properties consistent with ACs 
found in other dSph galaxies in terms of their P-L and 
period-amplitude relations.  The other star, and the single supra-HB star in 
And~I, may be P2C variables rather than ACs.  Further observations are 
needed to clarify their nature though we note that in the Galactic globular 
cluster population, P2Cs are found only in those clusters with strong blue 
HB morphology, which both And~I and And~III lack.  Assuming that these 
two stars are ACs, we have included the resulting AC specific frequencies 
for And~I and And~III with the data for other dSph galaxies from Paper~II\@.  
These AC specific frequencies follow the trends defined by the other dSphs when 
plotted against both (visual) luminosity and mean abundance.  As we 
concluded in Paper~II, there is a clear correlation between the AC 
specific frequency and dSph luminosity (and mean abundance) as first 
noted by Mateo et al.\ (1995).  

For the RRab stars in And~I, we found $\langle {\rm [Fe/H]} \rangle = -1.49$ 
from the Alcock et al.\ (2000) equation that relates the metallicity of a 
RRab star to its period and amplitude.  This mean abundance estimate agrees 
well with that of DACS00, $\langle {\rm [Fe/H]} \rangle = -1.46\pm0.12$, 
determined from the colors of the red giant branch stars.  For And~III, 
applying the same method yields $\langle {\rm [Fe/H]} \rangle = -1.81$ 
from the RRab stars, which again agrees well with the mean abundance from the red 
giant branch, $\langle {\rm [Fe/H]} \rangle = -1.88\pm0.11$ (DAC02).  
Indeed, we have found that the Alcock et al.\ (2000) equation provides 
a good estimate for the mean metallicity for all the four systems we have 
surveyed despite the fact that not all red giant branch stars evolve to become 
RRL stars.  It remains to be seen how well this equation applies to other 
dwarf galaxies or globular clusters.

In agreement with Baldacci et al.\ (2003), Marconi et al.\ (2004), and 
Gallart et al.\ (2004), we find that the short-period Cepheids in dwarf 
galaxies such as Leo~A, Sextans~A, and IC~1613 do not follow the same P-L 
relations as found for the ACs.  The short-period Cepheids 
typically have higher luminosities than the ACs at fixed period.

In summary, our search for variable stars in the M31 dSph companions has, 
up to this point, revealed that their variable star populations are basically 
indistinguishable from those of their Galactic counterparts.  In particular, 
the properties and specific frequencies of the 
ACs in the M31 dSph galaxies are similar to those found in the Galactic 
dSph galaxies.  Further, the properties of the RRL stars are consistent with 
those found in the Galactic dSphs, and have been shown to be good indicators 
for the distance and, perhaps surprisingly, also the mean metallicity for 
these M31 dSph companions.

\acknowledgements

This research was supported in part by NASA through grant number GO-08272 
from the Space Telescope Science Institute, which is operated by AURA, Inc., 
under NASA contract NAS 5-26555.

Thank you to the referee for the valuable comments for this paper.  
We would like to thank P. B. Stetson for graciously sharing his PSFs for 
the WFPC2 and for the use of his data reduction programs.  Thanks to A. C. 
Layden for the use of his light curve analysis programs.  GDaC is 
grateful to Andrew Drake for his efforts in carrying out the initial 
analysis of the variable stars in And~I\@.  Thanks also to M. Catelan and 
H. A. Smith for discussions on the RRL period distribution in And~III.

\clearpage 

\begin{deluxetable*}{cccc} 
\tablewidth{0pc} 
\tablecaption{Photometric Differences \label{tbl-1}} 
\tablehead{
& \colhead{Chip} & \colhead{$V$} & \colhead{$B$} 
          } 
\startdata 
And~I:   & WFC2 & -0.07 & -0.10 \\
         & WFC3 & -0.05 & -0.09 \\ 
         & WFC4 & -0.07 & -0.08 \\ 
And~III: & PC   & -0.05 & -0.08 \\ 
         & WFC2 & -0.02 & -0.05 \\ 
         & WFC3 & -0.03 & -0.05 \\
         & WFC4 & -0.01 & -0.03 \\
\enddata
\tablecomments{difference = magnitude in present study - CMD magnitude} 
\end{deluxetable*} 

\LongTables 
\begin{deluxetable*}{cccccccccccc} 
\tablewidth{0pc} 
\tabletypesize{\scriptsize} 
\tablecaption{Light Curve Properties for And~I \label{tbl-2}} 
\tablehead{
\colhead{ID} & \colhead{RA (2000)} & \colhead{Dec (2000)} & \colhead{Period} & 
\colhead{$\langle V \rangle$} & 
\colhead{$\langle B \rangle$} & \colhead{$(B-V)_{\rm mag}$}  & 
\colhead{$A_V$} & \colhead{$A_B$} & \colhead{$\Delta\log\,P$} & 
\colhead{[Fe/H]} & \colhead{Classification} \\ 
          } 
\startdata 
V01  & 0:45:42.18 & 38:02:51.8 & 1.630 & 24.336 & 24.665 & 0.338 & 0.39 & 0.55 & \nodata & \nodata & AC?, P2C? \\
V02  & 0:45:42.59 & 38:02:43.9 & 0.348 & 24.802 & 25.153 & 0.366 & 0.50 & 0.71 & \nodata & \nodata & c \\
V03  & 0:45:42.57 & 38:03:00.4 & 0.412 & 25.084 & 25.404 & 0.334 & 0.48 & 0.68 & \nodata & \nodata & c; Drake: P=0.383 \\
V04  & 0:45:42.16 & 38:02:47.0 & \nodata & \nodata & \nodata & \nodata & \nodata & \nodata & \nodata & \nodata & RRL?; No F555W \\
V05  & 0:45:41.10 & 38:03:03.9 & 0.654 & 25.053 & 25.522 & 0.473 & 0.27 & 0.39 &  0.05 & -1.33 & ab \\
V06  & 0:45:42.63 & 38:03:14.1 & 0.430 & 25.010 & 24.434 & 0.433 & 0.38 & 0.54 & \nodata & \nodata & c \\
V07  & 0:45:40.30 & 38:03:47.2 & 0.338 & 25.137 & 25.528 & 0.408 & 0.55 & 0.77 & \nodata & \nodata & c; Drake: P=0.358 \\
V08  & 0:45:39:27 & 38:03:33.0 & 0.376 & 25.019 & 25.502 & 0.493 & 0.43 & 0.61 & \nodata & \nodata & c \\
V09  & 0:45:43.68 & 38:03:49.0 & 0.413 & 25.139 & 25.567 & 0.447 & 0.57 & 0.81 & \nodata & \nodata & c \\
V10  & 0:45:44.89 & 38:03:24.7 & 0.664 & 25.048 & 25.442 & 0.423 & 0.83 & 1.17 & -0.06 & -2.13 & ab; Drake: P=0.667 \\
V11  & 0:45:41.64 & 38:03:18.0 & 0.349 & 25.226 & 25.568 & 0.349 & 0.35 & 0.49 & \nodata & \nodata & c \\
V12  & 0:45:42.17 & 38:03:47.3 & 0.367 & 25.145 & 25.436 & 0.303 & 0.46 & 0.65 & \nodata & \nodata & c; Drake: P=0.323 \\
V13  & 0:45:42.30 & 38:02:59.6 & 0.716 & 25.082 & 25.459 & 0.395 & 0.67 & 0.95 & -0.07 & -2.21 & ab \\
V14  & 0:45:38.82 & 38:03:08.0 & 0.471 & 25.145 & 25.533 & 0.394 & 0.31 & 0.44 & \nodata & \nodata & c; Drake: P=0.367 \\
V15  & 0:45:40.89 & 38:03:00.2 & 0.355 & 25.187 & 25.561 & 0.383 & 0.42 & 0.59 & \nodata & \nodata & c; Drake: P=0.355 \\
V16  & 0:45:43.53 & 38:03:32.9 & 0.665 & 25.061 & 25.510 & 0.472 & 0.74 & 1.05 & -0.05 & -2.02 & ab \\
V17  & 0:45:43.32 & 38:02:53.6 & 0.350 & 25.265 & 25.663 & 0.404 & 0.39 & 0.56 & \nodata & \nodata & c \\
V18  & 0:45:40.20 & 38:03:23.0 & 0.529 & 25.126 & 25.600 & 0.500 & 0.78 & 1.10 &  0.05 & -1.19 & ab; Drake: P=0.479 \\
V19  & 0:45:42.91 & 38:03:39.3 & 0.458 & 25.218 & 25.432 & 0.278 & 1.19 & 1.70 &  0.03 & -1.18 & ab \\
V20  & 0:45:41.04 & 38:03:58.0 & 0.513 & 25.254 & 25.583 & 0.337 & 0.42 & 0.60 & \nodata & \nodata & c; Drake: P=0.509 \\
V21  & 0:45:43.64 & 38:03:26.7 & 0.470 & 25.323 & 25.703 & 0.385 & 0.29 & 0.41 & \nodata & \nodata & c \\
V22  & 0:45:43.70 & 38:03:04.4 & 0.484 & 25.173 & 25.538 & 0.420 & 1.13 & 1.60 &  0.02 & -1.31 & ab \\
V23  & 0:45:41.91 & 38:03:31.0 & 0.471 & 25.358 & 25.692 & 0.340 & 0.37 & 0.52 & \nodata & \nodata & c \\
V24  & 0:45:42.91 & 38:02:38.5 & 0.491 & 25.052 & 25.504 & 0.484 & 0.85 & 1.20 &  0.07 & -1.00 & ab; Drake: P=0.491 \\
V25  & 0:45:40.91 & 38:03:25.0 & 0.615 & 25.206 & 25.661 & 0.471 & 0.58 & 0.82 &  0.02 & -1.50 & ab \\
V26  & 0:45:43.83 & 38:03:24.4 & 0.515 & 25.157 & 25.598 & 0.477 & 0.90 & 1.27 &  0.04 & -1.25 & ab \\
V27  & 0:45:42.96 & 38:03:19.3 & 0.577 & 25.081 & 25.515 & 0.473 & 0.97 & 1.36 & -0.03 & -1.77 & ab \\
V28  & 0:45:41.56 & 38:03:09.4 & 0.510 & 25.089 & 25.517 & 0.474 & 1.04 & 1.47 &  0.02 & -1.39 & ab; Drake: P=0.567 \\
V29  & 0:45:40.02 & 38:03:07.7 & 0.498 & 25.173 & 25.530 & 0.401 & 0.95 & 1.35 &  0.04 & -1.18 & ab \\
V30  & 0:45:42.55 & 38:03:28.7 & 0.381 & 25.212 & 25.642 & 0.450 & 0.60 & 0.80 & \nodata & \nodata & c \\
V31  & 0:45:45.14 & 38:03:22.6 & 0.494 & 25.032 & 25.408 & 0.433 & 1.15 & 1.62 &  0.01 & -1.41 & ab \\
V32  & 0:45:42.84 & 38:03:28.4 & 0.582 & 25.196 & 25.663 & 0.482 & 0.61 & 0.86 &  0.04 & -1.33 & ab \\
V33  & 0:45:42.73 & 38:03:23.1 & 0.537 & 25.220 & 25.652 & 0.473 & 0.92 & 1.29 &  0.02 & -1.42 & ab; Drake: P=0.537 \\
V34  & 0:45:42.64 & 38:02:55.6 & 0.510 & 25.224 & 25.596 & 0.405 & 0.86 & 1.22 &  0.05 & -1.15 & ab \\
V35  & 0:45:38.84 & 38:02:59.0 & 0.637 & 25.359 & 25.755 & 0.408 & 0.51 & 0.72 &  0.02 & -1.54 & ab \\
V36  & 0:45:41.32 & 38:03:15.9 & 0.575 & 25.180 & 25.629 & 0.476 & 0.76 & 1.07 &  0.01 & -1.48 & ab \\
V37  & 0:45:42.75 & 38:02:42.3 & 0.534 & 25.184 & 25.568 & 0.434 & 1.09 & 1.54 & -0.01 & -1.63 & ab; Drake: P=0.674 \\
V38  & 0:45:42.49 & 38:03:34.9 & 0.520 & 25.199 & 25.648 & 0.480 & 0.79 & 1.11 &  0.05 & -1.13 & ab \\
V39  & 0:45:38.45 & 38:03:18.2 & 0.517 & 25.161 & 25.578 & 0.480 & 1.21 & 1.70 & -0.02 & -1.66 & ab; Drake: P=0.569 \\
V40  & 0:45:41.62 & 38:03:17.8 & 0.522 & 25.231 & 25.676 & 0.474 & 0.79 & 1.12 &  0.05 & -1.15 & ab \\
V41  & 0:45:41.47 & 38:02:57.4 & 0.484 & 25.254 & 25.616 & 0.419 & 1.05 & 1.48 &  0.04 & -1.20 & ab; Drake: P=0.591 \\
V42  & 0:45:40.04 & 38:03:29.4 & 0.447 & 25.282 & 25.686 & 0.465 & 1.08 & 1.53 &  0.07 & -0.94 & ab \\
V43  & 0:45:39.62 & 38:03:42.3 & 0.746 & 25.276 & 25.650 & 0.383 & 0.44 & 0.62 & -0.04 & -2.06 & ab \\
V44  & 0:45:43.99 & 38:03:10.9 & 0.748 & 25.169 & 25.552 & 0.386 & 0.24 & 0.33 & -0.01 & -1.80 & ab \\
V45  & 0:45:49.15 & 38:02:02.4 & 0.772 & 24.988 & 25.493 & 0.508 & 0.29 & 0.41 & -0.03 & -1.99 & ab \\
V46  & 0:45:45.02 & 38:02:55.0 & 0.589 & 25.039 & 25.456 & 0.438 & 0.64 & 0.91 &  0.02 & -1.42 & ab \\
V47  & 0:45:47.03 & 38:03:14.7 & 0.400 & 25.043 & 25.411 & 0.377 & 0.39 & 0.55 & \nodata & \nodata & c; Drake: P=0.399 \\
V48  & 0:45:46.91 & 38:03:27.6 & 0.415 & 25.100 & 25.533 & 0.447 & 0.57 & 0.81 & \nodata & \nodata & c \\
V49  & 0:45:47.84 & 38:02:26.7 & 0.566 & 25.200 & 25.651 & 0.488 & 0.88 & 1.23 &  0.00 & -1.58 & ab; Drake: P=0.513 \\
V50  & 0:45:45.73 & 38:02:27.4 & 0.389 & 25.067 & 25.503 & 0.440 & 0.27 & 0.38 & \nodata & \nodata & c; Drake: P=0.335 \\
V51  & 0:45:48.87 & 38>02:37.8 & 0.363 & 25.118 & 25.652 & 0.546 & 0.44 & 0.62 & \nodata & \nodata & c; Drake: P=0.39 \\
V52  & 0:45:47.13 & 38:02:16.1 & 0.646 & 25.146 & 25.639 & 0.507 & 0.56 & 0.79 &  0.00 & -1.66 & ab; Drake: P=0.519 \\
V53  & 0:45:46.67 & 38:02:34.3 & 0.285 & 25.270 & 25.616 & 0.358 & 0.53 & 0.75 & \nodata & \nodata & c; Drake: P=0.522 \\
V54  & 0:45:50.71 & 38:02:42.8 & 0.597 & 24.932 & 25.330 & 0.436 & 0.87 & 1.23 & -0.02 & -1.77 & ab \\
V55  & 0:45:44.30 & 38:02:31.1 & 0.606 & 24.969 & 25.309 & 0.381 & 0.98 & 1.39 & -0.05 & -1.98 & ab \\
V56  & 0:45:46.20 & 38:03:17.5 & 0.628 & 25.119 & 25.602 & 0.508 & 0.73 & 1.03 & -0.02 & -1.78 & ab \\
V57  & 0:45:48.17 & 38:02:07.0 & 0.587 & 25.117 & 25.550 & 0.452 & 0.67 & 0.94 &  0.02 & -1.44 & ab; Drake: P=0.585 \\
V58  & 0:45:47.87 & 38:02:40.4 & 0.532 & 25.182 & 25.591 & 0.447 & 0.86 & 1.22 &  0.03 & -1.32 & ab; Drake: P=0.533 \\
V59  & 0:45:47.06 & 38:02:18.5 & 0.492 & 25.246 & 25.570 & 0.380 & 1.14 & 1.61 &  0.01 & -1.39 & ab; Drake: P=0.541 \\
V60  & 0:45:46.49 & 38:02:16.6 & 0.626 & 24.964 & 25.420 & 0.485 & 0.81 & 1.14 & -0.03 & -1.87 & ab \\
V61  & 0:45:45.37 & 38:02:27.7 & 0.554 & 25.192 & 25.651 & 0.474 & 0.60 & 0.85 &  0.06 & -1.13 & ab \\
V62  & 0:45:44.91 & 38:02:25.5 & 0.691 & 25.237 & 25.741 & 0.515 & 0.48 & 0.67 & -0.01 & -1.82 & ab; Drake: P=0.618 \\
V63  & 0:45:47.95 & 38:02:40.8 & 0.456 & 25.251 & 25.734 & 0.490 & 0.35 & 0.49 & \nodata & \nodata & c; Drake: P=0.388 \\
V64  & 0:45:45.35 & 38:02:30.6 & 0.564 & 25.178 & 25.649 & 0.490 & 0.67 & 0.95 &  0.04 & -1.29 & ab \\
V65  & 0:45:48.55 & 38:02:20.1 & 0.549 & 25.147 & 25.585 & 0.490 & 1.11 & 1.56 & -0.03 & -1.77 & ab; Drake: P=0.499 \\
V66  & 0:45:46.07 & 38:03:11.0 & 0.514 & 25.183 & 25.617 & 0.479 & 0.95 & 1.33 &  0.03 & -1.30 & ab; Drake: P=0.513 \\
V67  & 0:45:43.99 & 38:02:25.2 & 0.363 & 25.346 & 25.745 & 0.406 & 0.35 & 0.50 & \nodata & \nodata & c \\
V68  & 0:45:45.44 & 38:02:30.4 & 0.555 & 25.222 & 25.657 & 0.456 & 0.69 & 0.97 &  0.04 & -1.25 & ab; Drake: P=0.496 \\
V69  & 0:45:48.95 & 38:03:15.0 & 0.512 & 25.175 & 25.591 & 0.448 & 0.87 & 1.23 &  0.04 & -1.18 & ab; Drake: P=0.514 \\
V70  & 0:45:45.84 & 38:02:34.0 & 0.608 & 25.205 & 25.703 & 0.512 & 0.56 & 0.79 &  0.03 & -1.43 & ab \\
V71  & 0:45:49.91 & 38:02:41.0 & 0.576 & 25.119 & 25.525 & 0.437 & 0.85 & 1.20 &  0.00 & -1.61 & ab \\
V72  & 0:45:44.42 & 38:02:42.9 & 0.441 & 25.135 & 25.517 & 0.440 & 1.08 & 1.52 &  0.07 & -0.89 & ab \\
V73  & 0:45:45.65 & 38:02:39.8 & 0.537 & 25.223 & 25.647 & 0.471 & 0.99 & 1.39 &  0.00 & -1.52 & ab \\
V74  & 0:45:49.21 & 38>02:30.9 & 0.535 & 25.248 & 25.568 & 0.380 & 1.07 & 1.51 & -0.01 & -1.62 & ab \\
V75  & 0:45:47.46 & 38:02:33.9 & 0.604 & 25.187 & 25.551 & 0.409 & 0.94 & 1.33 & -0.04 & -1.91 & ab; Drake: P=0.599 \\
V76  & 0:45:46.80 & 38:02:31.6 & 0.468 & 25.542 & 25.908 & 0.426 & 1.11 & 1.56 &  0.04 & -1.16 & ab \\
V77  & 0:45:47.50 & 38:03:01.1 & 0.310 & 25.240 & 25.626 & 0.392 & 0.31 & 0.43 & \nodata & \nodata & c; Drake: P=0.310 \\
V78  & 0:45:49.64 & 38:02:24.4 & 0.638 & 25.190 & 25.720 & 0.537 & 0.41 & 0.58 &  0.03 & -1.42 & ab; Drake: P=0.645 \\
V79  & 0:45:46.31 & 38:01:51.9 & 0.652 & 24.958 & 25.341 & 0.399 & 0.57 & 0.81 & -0.01 & -1.71 & ab \\
V80  & 0:45:41.50 & 38:01:34.0 & 0.356 & 24.999 & 25.441 & 0.456 & 0.50 & 0.71 & \nodata & \nodata & c; Drake: P=0.357 \\
V81  & 0:45:44.16 & 38:01:36.9 & 0.381 & 25.221 & 25.552 & 0.340 & 0.40 & 0.56 & \nodata & \nodata & c \\
V82  & 0:45:47.36 & 38:01:25.4 & 0.577 & 25.136 & 25.528 & 0.406 & 0.53 & 0.74 &  0.06 & -1.19 & ab; Drake: P=0.524 \\
V83  & 0:45:45.42 & 38:01:44.0 & 0.529 & 25.246 & 25.643 & 0.416 & 0.65 & 0.92 &  0.07 & -1.02 & ab; Drake: P=0.532 \\
V84  & 0:45:42.40 & 38:02:04.6 & 0.334 & 25.159 & 25.570 & 0.419 & 0.37 & 0.52 & \nodata & \nodata & c \\
V85  & 0:45:44.38 & 38:01:18.0 & 0.632 & 25.222 & 25.621 & 0.411 & 0.55 & 0.78 &  0.01 & -1.57 & ab; Drake: P=0.707 \\
V86  & 0:45:44.02 & 38:01:03.6 & 0.549 & 25.144 & 25.651 & 0.527 & 0.64 & 0.89 &  0.06 & -1.14 & ab \\
V87  & 0:45:42.42 & 38:01:22.3 & 0.669 & 25.085 & 25.491 & 0.439 & 0.88 & 1.25 & -0.08 & -2.22 & ab \\
V88  & 0:45:43.01 & 38:02:17.0 & 0.565 & 24.996 & 25.402 & 0.452 & 1.03 & 1.45 & -0.03 & -1.77 & ab \\
V89  & 0:45:44.90 & 38:02:09.8 & 0.523 & 25.180 & 25.467 & 0.335 & 1.05 & 1.48 &  0.00 & -1.50 & ab \\
V90  & 0:45:41.86 & 38:01:36.3 & 0.610 & 25.098 & 25.526 & 0.448 & 0.67 & 0.94 &  0.01 & -1.59 & ab \\
V91  & 0:45:42.03 & 38:01:10.3 & 0.594 & 25.125 & 25.549 & 0.453 & 0.78 & 1.10 &  0.00 & -1.63 & ab \\
V92  & 0:45:43.13 & 38:01:24.0 & 0.469 & 25.062 & 25.381 & 0.379 & 1.07 & 1.52 &  0.05 & -1.11 & ab; Drake: P=0.513 \\
V93  & 0:45:45.42 & 38:00:55.1 & 0.476 & 25.213 & 25.510 & 0.351 & 1.04 & 1.48 &  0.04 & -1.13 & ab \\
V94  & 0:45:42.29 & 38:02:00.6 & 0.605 & 25.258 & 25.747 & 0.530 & 0.93 & 1.31 & -0.04 & -1.90 & ab \\
V95  & 0:45:46.05 & 38:01:16.2 & 0.457 & 25.158 & 25.536 & 0.443 & 1.22 & 1.72 &  0.03 & -1.21 & ab; Drake: P=0.499 \\
V96  & 0:45:46.20 & 38:01:00.1 & 0.520 & 25.271 & 25.711 & 0.479 & 0.89 & 1.25 &  0.04 & -1.27 & ab; Drake: P=0.578 \\
V97  & 0:45:43.30 & 38:01:58.8 & 0.603 & 25.136 & 25.602 & 0.488 & 0.73 & 1.03 &  0.00 & -1.62 & ab \\
V98  & 0:45:46.39 & 38:02:01.9 & 0.625 & 25.209 & 25.586 & 0.393 & 0.55 & 0.78 &  0.02 & -1.52 & ab \\
V99  & 0:45:45.62 & 38:01:13.5 & 0.716 & 25.215 & 25.670 & 0.468 & 0.51 & 0.71 & -0.04 & -1.99 & ab \\
V100 & 0:45:42.36 & 38:02:00.5 & 0.782 & 25.766 & 26.208 & 0.453 & 0.49 & 0.68 & -0.07 & -2.31 & ab \\
\enddata
\tablecomments{The periods referenced as Drake derive from the first year 
Ph.D. project (1996) of A. Drake with supervisor G. S. Da~Costa.  The 
published reference to this data is Da~Costa et al.\ (1997).}
\end{deluxetable*} 

\clearpage

\begin{deluxetable*}{cccccccccccc} 
\tablewidth{0pc} 
\tabletypesize{\scriptsize} 
\tablecaption{Light Curve Properties for And~III \label{tbl-3}} 
\tablehead{
\colhead{ID} & \colhead{RA (2000)} & \colhead{Dec (2000)} & \colhead{Period} & 
\colhead{$\langle V \rangle$} & 
\colhead{$\langle B \rangle$} & \colhead{$(B-V)_{\rm mag}$}  & 
\colhead{$A_V$} & \colhead{$A_B$} & \colhead{$\Delta\log\,P$} & 
\colhead{[Fe/H]} & \colhead{Classification} \\ 
          } 
\startdata 
V01 & 0:35:32.06 & 36:30:38.2 & 0.834 & 23.517 & 23.723 & 0.285 & 1.88 & 1.33 & \nodata & \nodata & AC \\
V02 & 0:35:31.74 & 36:30:23.8 & 0.590 & 25.013 & 25.287 & 0.307 & 0.80 & 1.13 & -0.01 & -1.63 & ab \\
V03 & 0:35:30.63 & 36:30:37.2 & 0.773 & 25.033 & 25.367 & 0.349 & 0.47 & 0.67 & -0.06 & -2.23 & ab \\
V04 & 0:35:31.22 & 36:30:21.1 & 0.629 & 25.031 & 25.308 & 0.313 & 0.91 & 1.29 & -0.05 & -2.03 & ab \\
V05 & 0:35:31.42 & 36:30:37.5 & 0.650 & 25.024 & 25.350 & 0.342 & 0.53 & 0.75 &  0.00 & -1.65 & ab \\
V06 & 0:35:36.28 & 36:29:30.4 & 0.678 & 23.166 & 23.575 & 0.433 & 0.93 & 0.66 & \nodata & \nodata & AC \\
V07 & 0:35:32.61 & 36:29:05.1 & 0.480 & 23.999 & 24.458 & 0.476 & 0.83 & 0.59 & \nodata & \nodata & AC \\
V08 & 0:35:33.01 & 36:29:14.7 & 1.51  & 24.303 & 24.821 & 0.534 & 0.89 & 0.63 & \nodata & \nodata & AC?, P2C? \\
V09 & 0:35:35.88 & 36:29:28.6 & 0.501 & 24.584 & 24.993 & 0.436 & 1.14 & 0.81 & \nodata & \nodata & AC \\
V10 & 0:35:32.51 & 36:30:05.0 & 0.678 & 25.037 & 25.322 & 0.305 & 0.67 & 0.95 & -0.04 & -2.00 & ab \\
V11 & 0:35:32.27 & 36:30:15.3 & 0.302 & 24.891 & 25.256 & 0.378 & 0.46 & 0.65 & \nodata & \nodata & c \\
V12 & 0:35:33.58 & 36:30:04.7 & 0.434 & 24.853 & 25.332 & 0.489 & 0.47 & 0.67 & \nodata & \nodata & c \\
V13 & 0:35:31.67 & 36:29:31.3 & 0.652 & 24.947 & 25.333 & 0.402 & 0.63 & 0.89 & -0.02 & -1.79 & ab \\
V14 & 0:35:34.29 & 36:29:17.1 & 0.666 & 25.093 & 25.328 & 0.442 & 0.74 & 1.04 & -0.05 & -2.02 & ab \\
V15 & 0:35:32.77 & 36:29:58.2 & 0.630 & 24.942 & 25.328 & 0.410 & 0.70 & 0.99 & -0.02 & -1.75 & ab \\
V16 & 0:35:34.93 & 36:29:43.9 & 0.620 & 25.054 & 25.295 & 0.266 & 0.76 & 1.08 & -0.02 & -1.77 & ab \\
V17 & 0:35:32.84 & 36:29:42.7 & 0.686 & 25.072 & 25.514 & 0.461 & 0.66 & 0.93 & -0.04 & -2.02 & ab \\
V18 & 0:35:36.01 & 36:29:25.4 & 0.613 & 24.972 & 25.293 & 0.345 & 0.66 & 0.93 &  0.01 & -1.59 & ab \\
V19 & 0:35:35.74 & 36:29:56.3 & 0.646 & 25.048 & 25.374 & 0.356 & 0.62 & 0.87 & -0.01 & -1.74 & ab \\
V20 & 0:35:31.16 & 36:29:04.1 & 0.661 & 24.977 & 25.425 & 0.468 & 0.66 & 0.93 & -0.03 & -1.88 & ab \\
V21 & 0:35:33.16 & 36:29:33.9 & 0.772 & 25.125 & 25.506 & 0.391 & 0.44 & 0.62 & -0.06 & -2.18 & ab \\
V22 & 0:35:36.68 & 36:29:09.9 & 0.455 & 24.968 & 25.352 & 0.407 & 0.68 & 0.96 & \nodata & \nodata & c \\
V23 & 0:35:32.94 & 36:29:59.1 & 0.616 & 25.003 & 25.343 & 0.355 & 0.58 & 0.83 &  0.02 & -1.51 & ab \\
V24 & 0:35:33.34 & 36:29:55.3 & 0.624 & 25.022 & 25.487 & 0.488 & 0.70 & 0.98 & -0.01 & -1.71 & ab \\
V25 & 0:35:27.61 & 36:29:45.6 & 0.761 & 24.892 & 25.351 & 0.464 & 0.32 & 0.44 & -0.03 & -1.97 & ab \\
V26 & 0:35:26.38 & 36:29:52.8 & 0.652 & 24.920 & 25.314 & 0.409 & 0.58 & 0.81 & -0.02 & -1.72 & ab \\
V27 & 0:35:26.61 & 36:29:59.5 & 0.491 & 25.091 & 25.443 & 0.365 & 0.48 & 0.67 & \nodata & \nodata & c \\
V28 & 0:35:29.88 & 36:29:33.9 & 0.702 & 25.069 & 25.495 & 0.447 & 0.66 & 0.92 & -0.05 & -2.11 & ab \\
V29 & 0:35:30.35 & 36:29:48.0 & 0.648 & 25.089 & 25.433 & 0.354 & 0.46 & 0.65 &  0.02 & -1.54 & ab \\
V30 & 0:35:30.42 & 36:29:47.0 & 0.740 & 25.117 & 25.431 & 0.326 & 0.51 & 0.72 & -0.05 & -2.11 & ab \\
V31 & 0:35:27.56 & 36:29:07.3 & 0.634 & 24.989 & 25.391 & 0.418 & 0.60 & 0.85 &  0.00 & -1.65 & ab \\
V32 & 0:35:28.94 & 36:29:20.0 & 0.617 & 25.070 & 25.388 & 0.349 & 0.86 & 1.22 & -0.04 & -1.88 & ab \\
V33 & 0:35:30.41 & 36:29:47.5 & 0.686 & 25.054 & 25.418 & 0.376 & 0.50 & 0.71 & -0.02 & -1.82 & ab \\
V34 & 0:35:30.70 & 36:30:12.3 & 0.617 & 25.039 & 25.350 & 0.332 & 0.71 & 1.00 & -0.01 & -1.68 & ab \\
V35 & 0:35:27.17 & 36:30:15.1 & 0.283 & 25.080 & 25.569 & 0.504 & 0.58 & 0.81 & \nodata & \nodata & c \\
V36 & 0:35:25.85 & 36:29:30.5 & 0.440 & 25.079 & 25.336 & 0.294 & 0.79 & 1.12 & \nodata & \nodata & c \\
V37 & 0:35:28.67 & 36:29:29.7 & 0.676 & 24.958 & 25.366 & 0.426 & 0.62 & 0.87 & -0.03 & -1.91 & ab \\
V38 & 0:35:30.11 & 36:31:12.1 & 0.714 & 24.932 & 25.265 & 0.348 & 0.58 & 0.82 & -0.05 & -2.07 & ab \\
V39 & 0:35:24.52 & 36:31:06.8 & 0.620 & 24.789 & 25.138 & 0.361 & 0.52 & 0.74 &  0.03 & -1.45 & ab \\
V40 & 0:35:27.07 & 36:30:45.9 & 0.407 & 24.933 & 25.253 & 0.339 & 0.57 & 0.81 & \nodata & \nodata & c \\
V41 & 0:35:28.70 & 36:30:17.4 & 0.671 & 24.983 & 25.386 & 0.415 & 0.51 & 0.72 & -0.01 & -1.74 & ab \\
V42 & 0:35:27.32 & 36:30:37.7 & 0.643 & 24.974 & 25.346 & 0.388 & 0.56 & 0.79 &  0.00 & -1.64 & ab \\
V43 & 0:35:29.74 & 36:30:36.0 & 0.362 & 25.088 & 25.397 & 0.323 & 0.48 & 0.68 & \nodata & \nodata & c \\
V44 & 0:35:27.70 & 36:31:02.4 & 0.537 & 24.914 & 25.265 & 0.394 & 0.93 & 1.31 &  0.01 & -1.44 & ab \\
V45 & 0:35:30.16 & 36:30:54.7 & 0.749 & 25.059 & 25.432 & 0.380 & 0.40 & 0.57 & -0.04 & -2.03 & ab \\
V46 & 0:35:29.22 & 36:30:27.6 & 0.353 & 25.061 & 25.351 & 0.308 & 0.56 & 0.79 & \nodata & \nodata & c \\
V47 & 0:35:29.09 & 36:30:32.7 & 0.657 & 25.037 & 25.372 & 0.353 & 0.67 & 0.94 & -0.03 & -1.87 & ab \\
V48 & 0:35:26.84 & 36:30:39.3 & 0.613 & 24.951 & 25.378 & 0.452 & 0.72 & 1.01 & -0.01 & -1.67 & ab \\
V49 & 0:35:28.16 & 36:30:15.2 & 0.720 & 25.010 & 25.432 & 0.434 & 0.49 & 0.70 & -0.04 & -1.99 & ab \\
V50 & 0:35:26.92 & 36:30:19.0 & 0.354 & 25.154 & 25.444 & 0.312 & 0.61 & 0.86 & \nodata & \nodata & c \\
V51 & 0:35:25.06 & 36:30:17.7 & 0.693 & 24.983 & 25.278 & 0.314 & 0.68 & 0.96 & -0.05 & -2.09 & ab \\
V52 & 0:35:27.25 & 36:30:17.2 & 0.447 & 25.116 & 25.407 & 0.301 & 0.42 & 0.59 & \nodata & \nodata & c \\
V53 & 0:35:29.07 & 36:30:22.1 & 0.534 & 25.085 & 25.387 & 0.356 & 1.04 & 1.48 & -0.01 & -1.57 & ab \\
V54 & 0:35:29.14 & 36:30:36.6 & 0.623 & 25.081 & 25.423 & 0.360 & 0.60 & 0.85 &  0.01 & -1.58 & ab \\
V55 & 0:35:28.25 & 36:30:49.1 & 0.599 & 25.013 & 25.294 & 0.308 & 0.80 & 1.13 & -0.01 & -1.69 & ab \\
V56 & 0:35:29.44 & 36:31:22.3 & 0.496 & 24.924 & 25.249 & 0.339 & 0.53 & 0.75 & \nodata & \nodata & c \\
\enddata
\end{deluxetable*} 

\clearpage

\begin{deluxetable*}{cccccc} 
\tablewidth{0pt} 
\footnotesize 
\tablecaption{$B$ Photometry of the Variable Stars in And~I \label{tbl-4}}
\tablehead{
\colhead{} & \multicolumn{2}{c}{V01} & & \multicolumn{2}{c}{V02} \\ 
\cline{2-3} \cline{5-6} \\ 
\colhead{HJD-2449000} & \colhead{$B$} & \colhead{$\sigma_{B}$} & & 
\colhead{$B$} & \colhead{$\sigma_{B}$} 
          }
\startdata 
575.837 &  24.937  &  0.089 &&  24.833  &  0.064 \\
575.904 &  24.779  &  0.108 &&  24.923  &  0.055 \\
575.971 &  24.896  &  0.118 &&  25.279  &  0.096 \\
576.038 &  24.920  &  0.140 &&  25.456  &  0.086 \\
576.105 &  24.754  &  0.081 &&  25.334  &  0.098 \\
576.172 &  24.818  &  0.142 &&  24.882  &  0.094 \\
581.266 &  24.424  &  0.055 &&  25.648  &  0.124 \\
581.333 &  24.442  &  0.092 &&  25.345  &  0.129 \\
581.400 &  24.325  &  0.060 &&  \nodata  &  \nodata \\
581.467 &  \nodata  &  \nodata &&  24.928  &  0.077 \\
581.534 &  24.366  &  0.080 &&  25.207  &  0.130 \\
581.601 &  24.450  &  0.061 &&  25.669  &  0.192 \\
\enddata 
\tablecomments{The complete version of this table is in the electronic 
edition of the Journal.  The printed edition contains only a sample.}
\end{deluxetable*}

\begin{deluxetable*}{cccccc} 
\tablewidth{0pt} 
\footnotesize 
\tablecaption{$V$ Photometry of the Variable Stars in And~I \label{tbl-5}}
\tablehead{
\colhead{} & \multicolumn{2}{c}{V01} & & \multicolumn{2}{c}{V02} \\ 
\cline{2-3} \cline{5-6} \\ 
\colhead{HJD-2449000} & \colhead{$V$} & \colhead{$\sigma_{V}$} & & 
\colhead{$V$} & \colhead{$\sigma_{V}$} 
          }
\startdata 
575.636 &  24.564  &  0.113 &&  24.928  &  0.084 \\
575.703 &  24.531  &  0.060 &&  25.126  &  0.076 \\
575.770 &  24.575  &  0.063 &&  24.862  &  0.093 \\
580.998 &  24.340  &  0.089 &&  24.734  &  0.081 \\
581.065 &  24.292  &  0.058 &&  24.576  &  0.075 \\
581.132 &  24.271  &  0.071 &&  24.734  &  0.042 \\
581.199 &  24.376  &  0.060 &&  24.856  &  0.069 \\
\enddata 
\tablecomments{The complete version of this table is in the electronic 
edition of the Journal.  The printed edition contains only a sample.}
\end{deluxetable*}

\begin{deluxetable*}{cccccc} 
\tablewidth{0pt} 
\footnotesize 
\tablecaption{$B$ Photometry of the Variable Stars in And~III \label{tbl-6}}
\tablehead{
\colhead{} & \multicolumn{2}{c}{V01} & & \multicolumn{2}{c}{V02} \\ 
\cline{2-3} \cline{5-6} \\ 
\colhead{HJD-2451000} & \colhead{$B$} & \colhead{$\sigma_{B}$} & & 
\colhead{$B$} & \colhead{$\sigma_{B}$} 
          }
\startdata 
231.743 &  \nodata  &  \nodata &&  25.439  &  0.217 \\
231.762 &  23.557  &  0.068 &&  25.499  &  0.197 \\
231.811 &  23.668  &  0.081 &&  25.303  &  0.250 \\
231.829 &  23.796  &  0.051 &&  25.436  &  0.194 \\
231.878 &  23.940  &  0.097 &&  24.448  &  0.141 \\
231.897 &  23.971  &  0.080 &&  24.751  &  0.130 \\
231.952 &  24.173  &  0.084 &&  24.872  &  0.097 \\
232.022 &  24.219  &  0.078 &&  25.141  &  0.125 \\
236.515 &  23.692  &  0.111 &&  25.762  &  0.261 \\
236.533 &  23.152  &  0.104 &&  \nodata  &  \nodata \\
236.582 &  22.807  &  0.080 &&  24.864  &  0.172 \\
236.601 &  22.924  &  0.117 &&  24.590  &  0.133 \\
236.649 &  22.962  &  0.111 &&  24.738  &  0.200 \\
236.667 &  23.195  &  0.116 &&  24.889  &  0.104 \\
236.716 &  23.356  &  0.118 &&  25.027  &  0.138 \\
236.735 &  23.439  &  0.113 &&  25.387  &  0.126 \\
\enddata 
\tablecomments{The complete version of this table is in the electronic 
edition of the Journal.  The printed edition contains only a sample.}
\end{deluxetable*} 

\clearpage

\begin{deluxetable*}{cccccc} 
\tablewidth{0pt} 
\footnotesize 
\tablecaption{$V$ Photometry of the Variable Stars in And~III \label{tbl-7}}
\tablehead{
\colhead{} & \multicolumn{2}{c}{V01} & & \multicolumn{2}{c}{V02} \\ 
\cline{2-3} \cline{5-6} \\ 
\colhead{HJD-2451000} & \colhead{$V$} & \colhead{$\sigma_{V}$} & & 
\colhead{$V$} & \colhead{$\sigma_{V}$} 
          }
\startdata 
231.613 &  22.916  &  0.033 &  25.265  &  0.203 \\
231.631 &  23.000  &  0.047 &  25.147  &  0.097 \\
231.675 &  23.131  &  0.032 &  25.325  &  0.165 \\
231.694 &  23.191  &  0.048 &  25.337  &  0.120 \\
236.381 &  23.967  &  0.073 &  25.262  &  0.137 \\
236.399 &  \nodata  &  \nodata &  25.341  &  0.144 \\
236.446 &  24.110  &  0.059 &  25.355  &  0.078 \\
236.464 &  24.116  &  0.065 &  25.326  &  0.096 \\
\enddata 
\tablecomments{The complete version of this table is in the electronic 
edition of the Journal.  The printed edition contains only a sample.}
\end{deluxetable*}

\begin{deluxetable*}{cccccc} 
\tablewidth{0pc} 
\footnotesize
\tablecaption{Absolute Magnitudes of the Supra-HB stars in And~I and And~III 
\label{tbl-8}}
\tablehead{
\colhead{System} & \colhead{ID} & \colhead{Mode} & \colhead{$M_V$} & 
\colhead{$M_B$} & \colhead{Classification} 
          }
\startdata 
And I   & V01 & \nodata & -0.31 & -0.03 & P2C or AC? \\ 
And III & V01 & H & -1.03 & -0.88 & AC \\ 
        & V06 & H & -1.38 & -1.03 & AC \\ 
        & V07 & H & -0.55 & -0.15 & AC \\ 
        & V08 & \nodata & -0.25 & 0.22 & P2C or AC? \\ 
        & V09 & F &  0.03 &  0.39 & AC \\
\enddata
\end{deluxetable*}

\begin{deluxetable*}{ccccccccccc} 
\tablewidth{0pt} 
\tabletypesize{\scriptsize} 
\tablecaption{Frequency of Anomalous Cepheids in Dwarf Spheroidal Galaxies
\label{tbl-9}}
\tablehead{
\colhead{Galaxy} & \colhead{$\langle {\rm [Fe/H]} \rangle$} & 
\colhead{$M_V$} & \colhead{PA} & \colhead{$1-b/a$} & 
\colhead{$r_{\rm exp}$} & \colhead{Surveyed $L$} & 
\colhead{$N_{\rm AC}$} & \colhead{$S$} & \colhead{$\log\,S$} & 
\colhead{References}
\\
 & & & (degrees) & & (arcsec) & (\%) & & & 
          } 
\startdata 
And I   & --1.5 & --11.7 &   0 & 0.0 & 1.40 & 17 & 1 & 0.15 & $-0.82^{+0.31}_{-{\rm inf}}$ & 1,2,3 \\
And III & --1.9 & --10.2 & 136 & 0.6 & 0.75 & 25 & 5 & 2.1  & $0.32^{+0.16}_{-0.24}$ & 1,2,4 \\ 
\enddata
\tablecomments{PA is the position angle measured from the North to the East.  
References:  (1) Caldwell et al.\ (1992); (2) This paper; (3) Da~Costa et al.\ 
(2000); (4) Da~Costa et al.\ (2002).  For the purposes of the discussion within the text 
we have assumed that And~I - V01 and And~III - V08 are anomalous Cepheids, but 
their true nature is still uncertain (see \S4).}
\end{deluxetable*}

\end{document}